\documentclass[10pt,aps,pra,twocolumn,showpacs,superscriptaddress,groupedaddress]{revtex4-1}  
\usepackage{amssymb,amsmath,graphicx}
\usepackage{graphicx,wrapfig,lipsum}
\usepackage[dvipsnames]{xcolor}
\usepackage{siunitx}
\usepackage{graphicx}  
\usepackage{dcolumn}   
\usepackage{bm}        
\usepackage{amssymb}   
\usepackage{braket}
\usepackage{mathtools}
\usepackage{hyperref}
\hypersetup{
    colorlinks,
    linkcolor={blue},
    citecolor={blue},
    urlcolor={blue}
}
\usepackage{stackengine}

\newcommand{\vect}[1]{\mathbf{#1}}

\usepackage{appendix}
\usepackage{subfigure}
\usepackage{ulem}
\newcommand{\bsf}[1]{\textsf{\textbf{#1}}}

\makeatletter
\def\@fnsymbol#1{}
\makeatother

\begin{document}
\title{Chemotactic Horizons Regulate Threshold Dynamics in the Biological Carbon Pump}

\author{Rahul Chajwa $^{1, \dagger}$, Manu Prakash $^{1,2,3,4, *}$.}
\thanks{* manup@stanford.edu \protect\\ $\dagger$ rchajwa@stanford.edu}
\affiliation{$^1$Department of Bioengineering, $^2$ Department of Biology, $^3$Department of Ocean, $^4$Woods Institute of the Environment, Stanford University, Stanford CA 94305 USA.}
\begin{abstract}
Microbial encounters with sinking marine snow regulate organic carbon remineralization, acting as a critical knob in the ocean’s biological carbon pump. Yet how the vast diversity of microbes in the open ocean respond to natural chemotactic cues coupling sedimentation-induced flow to microbial encounters in the ocean remains unresolved. Here we report shipboard measurements of microbial motility around freshly collected marine snow under imposed flows spanning ecologically relevant sinking speeds. Via high resolution single cell tracking at sea, we map a highly diverse repertoire of microbial motility, including rapid switching between motile and non-motile states. We identify a chemotactic horizon from measured reorientation towards chemical gradients, defining a spatial region around an aggregate within which chemotactic alignment overcomes vorticity-induced rotation due to sedimentation. This interplay produces a non-monotonic dependence of encounter rates on aggregate sinking speed, with a peak at 14.4 m day$^{-1}$ followed by a sharp decline beyond 78.1 m day$^{-1}$, with a tracer-like regime at high effective sinking speeds. We incorporate the measured encounter rates into a kinetic transport equation in position–velocity phase space, with an effective nonlinear damping arising from velocity-dependent remineralization. The resulting particle dynamics reveal a threshold-induced separatrix dividing overdamped and ballistic microbial populations, leading to velocity segregation. Our study connects direct measurements of microbial motility and sedimentation-induced flow to the emergent phase-space for fate of marine snow in the biological pump, establishing a quantitative multiscale framework for investigating ocean carbon export. 
\end{abstract}
\maketitle 
\section{Introduction}
Gravitational sinking of marine snow transports gigatons of carbon annually, from the surface to the deep ocean \cite{ALLDREDGE198841}. These particles sink through a microbially active ocean that harbours an extraordinarily diverse microbial ecosystem, with bacterial abundance typically $\mathcal{O}(10^6) \, \mathrm{cells/ mL}$ \cite{Azam1983Ecological, Azam2007}. Yet bacterial cells occupy a volume fraction of only $\mathcal{O}(10^{-7})$ in the ocean \cite{Ducklow2001b}, making direct microbe–marine snow encounters a key determinant of remineralization. Upon encounter, microbial degradation can dramatically alter aggregate morphology, density and chemical composition \cite{smith1992intense}, releasing dissolved compounds that are in-turn advected by sedimentation-induced viscous flows that can further attract more microbes \cite{kiorboe2008mechanistic, Seymour2010Chemoattraction}. The resulting dynamics of microbes in the ambient chemical and flow fields \cite{KiorboeJackson2001MarineSnow, Stocker2012}, shape the fate of sinking marine snow \cite{brumley2020cutting}. Conceptually disentangling these coupled microscale processes in the ocean remains a central issue in mechanistically linking microbial activity to the transport efficiency of the biological carbon pump \cite{Boyd2019, Ducklow2001}, critical for ocean scale modeling. 

As marine snow aggregates sink through the water column, individual microbes continually arrive and depart. This drives the assembly and succession of distinct aggregate-associated communities that perform degradation \cite{Stephens2024DirectObservations, Pelve2017BacterialSuccession, Durkin_2026}. Previous studies of swimming bacteria around sinking particles have derived encounter kernels to explain colonization rates \cite{Kiorboe2002Colonization, kiorboe2008mechanistic, Slomka2020}, while research on motility in complex environments has established motility–flow coupling \cite{Pedley1992, Stark2016, Durham2013} and chemotaxis \cite{blackburn1998microscale, berg2004ecoli, Locsei2009, barbara2003bacterial, Stocker2008, Stocker2012} as essential dynamical ingredients for colonization. 

While past work has elucidated several key mechanisms, understanding microbe–marine snow interactions in the open ocean has remained challenging \cite{Slomka2023EncounterRates, Turczynowicz_2026}. Experimental studies of microbial motility and colonization have often relied on artificial aggregates or selected lab cultured bacterial strains, while theoretical models of microbial motility have drawn heavily on chemotactic behavior of E. coli \cite{Locsei2009RunAT}. Standard culturing methods capture less than 1\% of environmental microbial diversity \cite{Ferguson1984}, presenting a severe sampling bias \cite{StaleyKonopka1985}. In contrast, motility in natural marine bacterial assemblages exhibits substantial variability in both the fraction of motile cells \cite{grossart2001bacterial, Keegstra2025} and their swimming gaits \cite{Mitchell1995NaturalAssemblages, johansen2002variability}. Furthermore, measured swimming speeds and acceleration in natural assemblages can differ markedly from those of cultured bacteria \cite{Mitchell1995NaturalAssemblages}. Adaptive and diverse microbial motility is a norm rather than an exception in the oceans \cite{grossart2001bacterial, luo2022diversity, johansen2002variability, Mitchell1995NaturalAssemblages}. This dynamics is coupled with the natural chemical and hydrodynamic fields generated by sinking aggregates, further complicating the prediction of microbial dynamics in natural settings \cite{Desai2019HydrodynamicColonization}. 
\begin{figure*}[t]
 \begin{center}
 \includegraphics[width=16.0 cm]{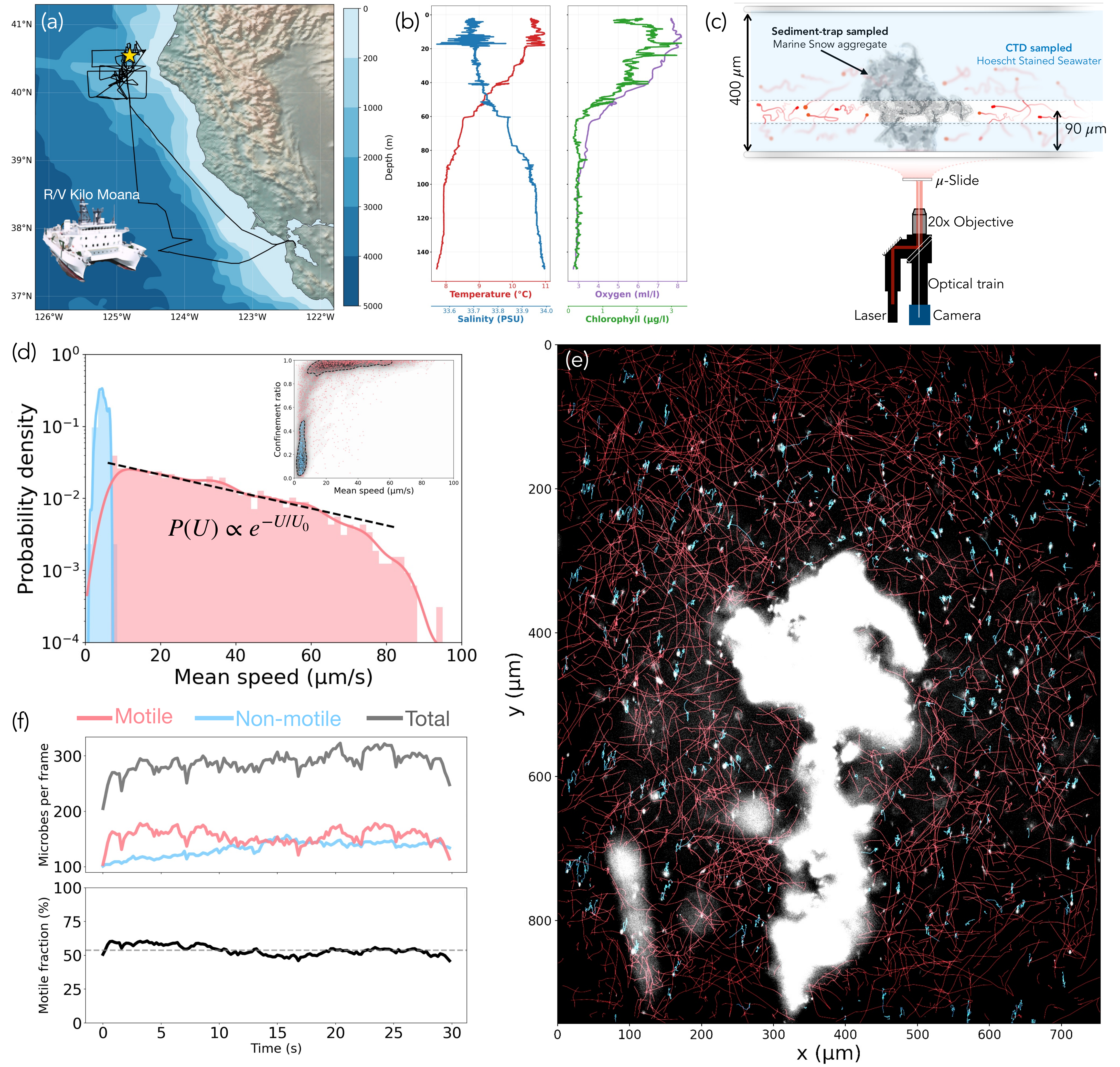}
 \caption{\label{Fig1} \textbf{Direct measurements of microbial motility around marine snow in open ocean} (a) We conducted imaging experiments aboard R/V Kilo Moana during an expedition VIISTA1 (Summer 2024) in California current ecosystem. Map shows expedition path with starred location $40.54^{\circ}$N, $124.81^{\circ}$W depicting the coordinates at which sampling and imaging were performed. (b) The CTD profiles of temperature, salinity, chlorophyll, and oxygen. Deep chlorophyll maximum (DCM) is at 20 m below surface. (c) Direct measurements of microbial community surrounding a marine snow particle on a table-top imaging setup at sea, with fresh marine snow aggregate collected from sediment trap at 80m placed in Ibidi micro-channel (depth = $400 \mathrm{\mu m}$) filled with Niskin bottle sampled seawater from DCM. Experimental setup utilizes an open source microscopy platform (SQUID \cite{Li2020Squid} in epifluorescence modality) to image quasi-two-dimensional trajectories of individual microbes swimming around mairne snow, using live nucleic acid stain (Hoescht). (d) Distribution of track-wise mean speed $U\equiv \langle v \rangle$ computed by tracking individual organisms. The kernel density estimation in the confinement Vs mean speed plane, shown in the inset, demarcates the non-motile (blue) and motile (red) sub-populations. The probability density is exponentially distributed ($\propto e^{-\mathrm{\bar{v}/v_0}}$) across a wide range of mean speed 7-70 $\mu$m/s, with $\mathrm{v_0} = 37.2 \mu$m/s ($R^{2} = 0.938)$.
 (e) Overlay of individual bacterial tracks around a stationary marine snow with non-motile (blue) and motile (red) microbes.
 (f) Total number of Motile and non-motile microbes as a function of time shows that their relative fraction fluctuates around $53.8\%$.
 }
 \label{Figure1}
 \end{center}
 \end{figure*}

Large-scale metagenomic and metatranscriptomic studies in the open ocean have mapped potential gene families to microbial traits like motility (flagellar genes), chemotaxis (sensory pathways and sensory repertoire) and receptor diversity \cite{Federico2009, Giovannoni2005, Hallstrom2022} across field-collected  microbial samples representing many uncultivated marine bacteria \cite{Tara2015}. Comparing freely suspensed and particle-attached microbial communities further highlight their diversification across niches \cite{luo2022diversity}. 
Genomic profiling can only indicate the potential phenotypes but cannot reveal the real-time dynamical behaviors of microbes that govern their interaction with the sinking marine snow. 
Microfluidic approaches have begun to elucidate chemotactic responses to dissolved chemical compounds in coastal microbial assemblages \cite{Lambert2017ISCA, Raina2022}; however, the dynamics of individual microbes around freshly collected marine snow under sedimentation-relevant flows in the open ocean have never been directly observed. 

Here, we approach this observational \textit{terra incognita} by directly measuring individual microbial trajectories around freshly collected marine snow. The experiments expose freshly collected marine snow aggregates collected by sediment traps (80m depth) to a natural microbial assemblage collected from the deep chlorophyll maximum (DCM) using a niskin bottle, with imaging commencing within 30 minutes of sample collection to ensure a close experimental proxy for microbe–marine snow interactions under native conditions. All experiments presented here are conducted aboard a research vessel in the open ocean using a dedicated tabletop setup (Fig 1). Throughout this study, \textit{microbes} refer to the DNA-stained cells individually tracked in our experiments, whose fluorescence-measured dimensions ranged from 0.25 to 2.4 $\mu$m. Many marine bacteria belong to this size range; however, the presence of some swimming nano-flagellates in our sample cannot be ruled out. \textit{Encounters} refers specifically to encounters between free-living suspended microbes and marine snow aggregates. Recorded single cell trajectories capture the projected two-dimensional motion of freely suspended microbes in the focal region, approximately $90 \mu\mathrm{m}$ from the transverse channel wall, where hydrodynamic wall effects are negligible.

We measure microbial dynamics around marine snow under controlled flows spanning ecologically relevant sinking speeds. Single cell tracking reveals a diverse repertoire of motility gaits, including  ballistic, run–stop–run, and run–reverse–flick motion. Against the backdrop of a steady-state motile fraction \cite{grossart2001bacterial}, we also observe rapid transitions between motile and non-motile states. These behaviors span a near-continuous spectrum in the velocity covariance–confinement plane, revealing substantial heterogeneity in microbial motility, significantly departing from the run-and-tumble dynamics of E. coli. Combining instantaneous accelerations of strongly chemotactic microbes with measurements of the surrounding flow and computed oxygen gradients, we quantify chemotactic responses and identify a chemotactic horizon: a diffuse boundary around an aggregate within which chemotactic alignment overturns vorticity-induced rotation.

We find that increasing background flow non-monotonically modulates microbial encounters: coupling between motility, flow and chemotaxis enhances downstream accumulation, maximizing encounter rates at an equivalent sinking speed of  14.4 m/day, beyond which encounter rates decline sharply. Above a critical speed of 78.1 m/day, microbial dynamics approach an effectively tracer-like regime. This non-monotonic response contrasts with the monotonic increase in encounter rates predicted by the classic encounter kernels with Sherwood-number corrections \cite{Kiorboe2002Colonization}. Recent models incorporate aspects of motility, chemotaxis, and flow coupling \cite{Slomka2020, Borer2022PorousMS}, but behavioral heterogeneity of diverse microbes in the open ocean remains unaccounted for. Finally, we incorporate the measured encounter rates into a minimal 1D kinetic transport equation \cite{huang1987statistical} through a sinking-velocity-dependent remineralization rate, moving beyond the steady-state hypothesis \cite{Giering2017}. The resulting dynamics reveal a threshold-induced separatrix in position–velocity phase space, partitioning particles by initial sinking speed into upper-ocean retention and deep-ocean export. This framework links microscale measurements of microbial encounters to ocean-scale carbon sequestration. Our study provides an approach to empirically constructing and constraining far-from-equilibrium models of ocean-carbon export that incorporate microbiological richness into sedimentation physics.

\begin{figure*}[t]
 \begin{center}
 \includegraphics[width=17.5 cm]{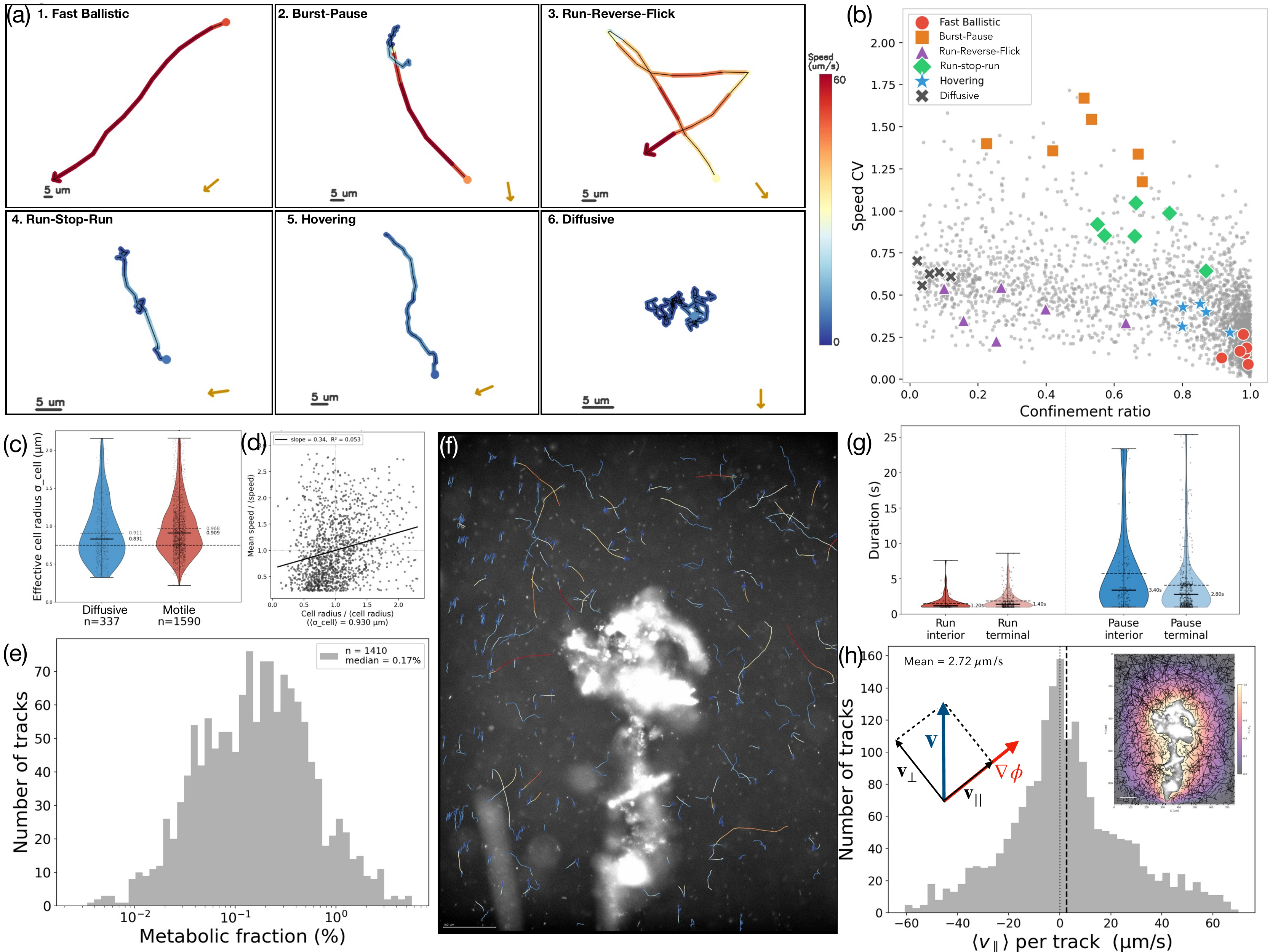}
 \caption{ \textbf{Motility repertoire and its statistics analyzed via comprehensive manual tracking of individual swimming bacteria across field collected dataset 
 (see methods for details):} 
(a) Subset of representative tracks for manually identified single cells annotated as dominant swimming categories. For each category a $5 \mu$m scale bar is on the bottom left and an arrow on bottom right points towards the aggregate. Colorbar depicts speed, such that speeds above 60 $\mu$/s is uniformly red. Richness of these swimming strategies highlight the importance of field based unbiased data collection. (b) Scatter of speed coefficient of variation and confinement ratio separates the visually identified swimmer categories. (c) The distribution of point spread function corrected size for non-motile (blue) and motile (red) subpopulations. (d) The speed and sizes are weakly positively correlated with slope 0.34 (R$^{2} = 0.053$, Spearman $r=0.266$, $p<0.001$). (e) Calculating the Stokesian dissipation or power from size and speeds is used to convert measured power into motility fraction using known flagellar power consumption and efficiency. (f) Run-and-pause dynamics shows rapid transition between non-motile and motile states that occurs uniformly around the aggregate, and (g) shows the corresponding distribution of run and pause times, with and without including terminal tracks. (h) Distribution of projection of the instantaneous velocity along gradient of the diffusive scalar $\nabla c$ shows weak but statistically significant chemotaxis towards the aggregate, with mean speed 2.72 $\mu$/s.}
 \label{Figure2}
 \end{center}
 \end{figure*}
 
 \section{Ocean-based Experiments in California Current Ecosystem}
Ocean campaigns have traditionally characterized microbial abundance, activity, and ecology using bulk measurements of microbial biomass and metabolic rates, alongside molecular profiling of microbial communities \cite{Kiorboe2003, ploug1999photosynthesis, MunsonMcGee2022, durkin2022tracing}. These methods probe genetic information or physicochemical signatures averaged across bulk seawater samples, leaving unresolved critical physiology of living microbes such as motile behavior around marine snow. Microbial motility has been studied extensively under controlled laboratory conditions for select bacteria\cite{Azam1983Ecological, Mitchell1995NaturalAssemblages, barbara2003bacterial, grossart2001bacterial}. But technical challenges associated with high-resolution imaging and single-cell tracking aboard research vessels have restricted such measurements primarily to coastal waters, as noted in \cite{Keegstra2025}.

We conducted shipboard experiments aboard the research vessel R/V Kilo Moana in the California Current System during cruise KM2410, as part of the VIISTA campaign (Virus Infection at the Interface of Sinking, Turbulence, and Aggregation) in Summer 2024. Our benchtop setup integrated seawater and marine snow collection at $40.54^{\circ}$N, $124.81^{\circ}$W with high-resolution fluorescence imaging in a microchannel [see Fig. \ref{Figure1}]. Seawater was collected from the deep chlorophyll maximum (DCM; 20 m depth) using a CTD rosette equipped with Niskin bottles. Marine snow aggregates were collected at 80 m depth using a freely suspended sediment trap (mesh size, $50\mathrm{\mu m}$; diameter, $2\mathrm{m}$) deployed in Lagrangian mode for 24 h \cite{Ben2015, Peterson2005}. The collected aggregates were sub-sampled using a quantitative splitter designed to minimize fluid shear \cite{Lamborg2008Flux}.
 
Seawater sample (45 mL) was stained with 0.89 $\mu$M Hoechst 33342 (Thermo Scientific, Cat. No. PI62249; 20 mM stock solution) and incubated for 30 min at \SI{4}{\degree}C. This fluorescent dye binds to A-T-rich regions of double-stranded DNA. To simulate sedimentation-induced flow, we placed an aggregate in an ibidi $\mu$-Slide I Luer (part no. 80176; channel dimensions, $5,\mathrm{mm} \times 50,\mathrm{mm} \times 0.4,\mathrm{mm}$) and pumped the live stained seawater through the channel using a peristaltic pump. The imposed flow speed, mimicking sinking marine snow particles, ranged from 0 to $820,\mathrm{\mu m,s^{-1}}$. The residual flow fluctuations introduced by peristaltic motion (Supplementary Fig. \ref{FigureS2}) were accounted for in our analysis through a time-averaging (see methods section C and supplementary section I.B).

Measurements were performed using SQUID, an open-source, modular microscopy platform \cite{Li2020Squid} that enables high-resolution fluorescence imaging at sea with a minimal, expedition-ready configuration that can be tradionally challenging in open ocean shipboard conditions. Stained microbes flowing through the microchannel were excited with a 355 nm laser and images were acquired at 5 frames s$^{-1}$ (fps) with a spatial resolution of 0.33 $\mu$m per pixel [Fig. \ref{Figure1}(c)]. Our image sampling frequency sits above the reported Nyquist limit of microbial reorientation ($ 0.5 - 3$ Hz) \cite{Stocker2016} but completely aliases the flagellar beating dynamics. For each imposed flow speed, we acquired a 2 min movie and analyzed a 30 s segment, balancing observation duration against data quality given the need for frame-by-frame manual error correction in tracking the individual microbe. We use the term “microbes” throughout because Hoechst staining labels DNA but does not establish taxonomic identity. Although cell size and swimming dynamics are consistent with motile and nonmotile marine prokaryotes, our assay is strictly phenotypic and does not identify individual organisms taxonomically.

\section{Direct Measurement of Microbial Motility around Marine Snow}
We optimize the staining and imaging parameters, and microscope configuration at sea to acquire single cell resolution while keeping the field of view large enough to capture the marine snow aggregate and the fluid region around it. Individual microbial tracks were captured at 5 fps with 200 millisecond exposure which optimizes fluorescence signal at the cost of time resolution and blurring the cell signal of motile tracks along the swimming velocity. The notorious experimental challenge of getting individual microbial tracks was solved using a custom analysis pipeline that combines algorithmic cell detection and trajectory tracking with frame-by-frame manual review and error correction of each track [see Methods]. This resulted in a total of 2657 tracks with $\geq 4$ spots. 

For the resulting tracks we calculate the confinement ratio (CR), defined as the ratio of net displacement to total distance traveled and mean speed ($\mathrm{\bar{v}}$), and fit a 2D gaussian kernel density estimate (KDE) in the CR-$\mathrm{\bar{v}}$ plane [Fig. \ref{Figure1} (d) inset]. The 20th-percentile density contour of the KDE evaluated over the sub-region (speed $<$ 10 $\mu$m/s, CR $<$ 0.5) is used as a bounding isoline: every track that falls inside that contour is identified as non-motile (N = 489, 18.4\%; blue color); everything else is motile (N=2168, 81.6\%; red color) [Fig. \ref{Figure1} (d)]. 1D KDE fitted on the probability density of mean speed for the swimming and non-swimming populations cross at 7.01 $\mu$m/s demarcates the motile and non-motile cells. Non-motile distribution departs from the Dirac delta function due to a combination of thermal diffusivity and mechanical vibration and pseudo forces from the ship's motion \cite{chajwa2024hidden} [see supplementary material for characterization of the ship background noise]. We find that the track-wise mean speed is exponentially distributed ($\sim e^{\mathrm{\bar{v}/v}_0}$, with $\mathrm{v_0 = 37.2 \mu}$m/s) over a wide range $7-70 \mu$m/s, declining sharply beyond $\simeq 70 \mu$m/s due to the tracking limitation imposed by low temporal resolution. All tracks overlayed show a nearly uniform distribution of cells, non-motile (blue) and motile (red), around the aggregate [see Fig. \ref{Figure1} (e)] with a bimodal motility statistics showing a steady state with mean $53.8\%$ motile within $30s$ time window [see Fig. \ref{Figure1} (f)].

Frame-by-frame inspection of each track of single microbial cells (example shown in Fig. \ref{Figure1} (e)) reveals highly diverse motility phenotypes. Here we highlight a few motility gaits that stand out to the human eye [see Fig. \ref{Figure2}(a)]: 1) ballistic: almost straight paths, including helical tracks with very small radius and large pitch, 2) run-to-pause and vice versa: reminiscent of `bursts of motility' noted once before in \cite{grossart2001bacterial}, presents an effective search strategy in low nutrient conditions, 3) run-reverse-flick: trajectories display sharp cusps, which are known to arise from recently studied flagellar buckling \cite{Son2013}, 4) run-stop-run: jerky runs with intermittent stops that last 0.5-3 seconds of non-motile phase which converts back to motile phase.   5) hovering: presents a continuous persistent random walk,  6) diffusive: non-motile state moving due to noise sources from both temperature and vibrating platform, with extremely low signal-to-noise ratio. The categories 1-6 differs markedly from the classic run-and-tumble motion often studied in common lab strains such as E.coli \cite{berg2004ecoli}. These traits segregate in the CR -speed coefficient of variance, albeit with a near continuous spectrum of motility behavior [see Fig. \ref{Figure2}(b)]. This diversity of behaviors represents the potential microbial species diversity present in each fluid parcel in the upper ocean.

 Next, using DNA fluorescence signal from each cell as an indirect proxy for cell size \cite{Gray2019}, we measure the point spread function corrected short dimension $b$ [see Methods], and observe that the median $b$ of non-motile cells is 0.831 $\mu$m, only slightly smaller than that of motile cells 0.909 $\mu$m, while $b$ and $\bar{\mathrm{v}}$ show weak positive correlation (Spearman r = 0.266, $p<0.001$), contrary to the theoretically expected negative correlation in the limit of small cell sizes \cite{Mitchell1991CellSize}. From the speed and corresponding size we estimate the metabolic fraction of energy invested in swimming based on: viscous power dissipation $6 \pi \mu b \bar{v}^2$ (assuming spherical cells), flagellar motor efficiency of $1.3 \%$ \cite{purcell1977life}, and the mass specific metabolic rate $0.2 \mathrm{pW/\mu m^{3}}$ \cite{makarieva2008mean} (multiplied by cell volume) [see Fig. \ref{Figure2}(e)], which gives a median metabolic fraction of $0.17 \%$ with maximum $\simeq 10\%$. To the best of our knowledge, this is the first at sea measurement of the metabolic cost of swimming microbes in the open ocean. Single cell tracking in field conditions allows us to comprehensively demonstrate a burst motility, in which a swimming microbe pauses for a period of time and again restarts motility [Fig. \ref{Figure2}(f)]. This was first observed in coastal samples \cite{grossart2001bacterial} and verified in lab strains \cite{johansen2002variability}, but had not been previously seen in open water samples. The measured run and pause times show brief runs (1.2 s, median) and long pauses (3.4 s, median)[Fig. \ref{Figure2}(g)], with some pauses as long as 25 s, limited by the observation window (30 s). Such dynamic motile-non-motile transitions further highlight the need for single cell measurements compared to populations statistics that obscure the dynamic nature of the bistable motility states.
 
 
\begin{figure*}[t]
 \begin{center}
 \includegraphics[width=16.4 cm]{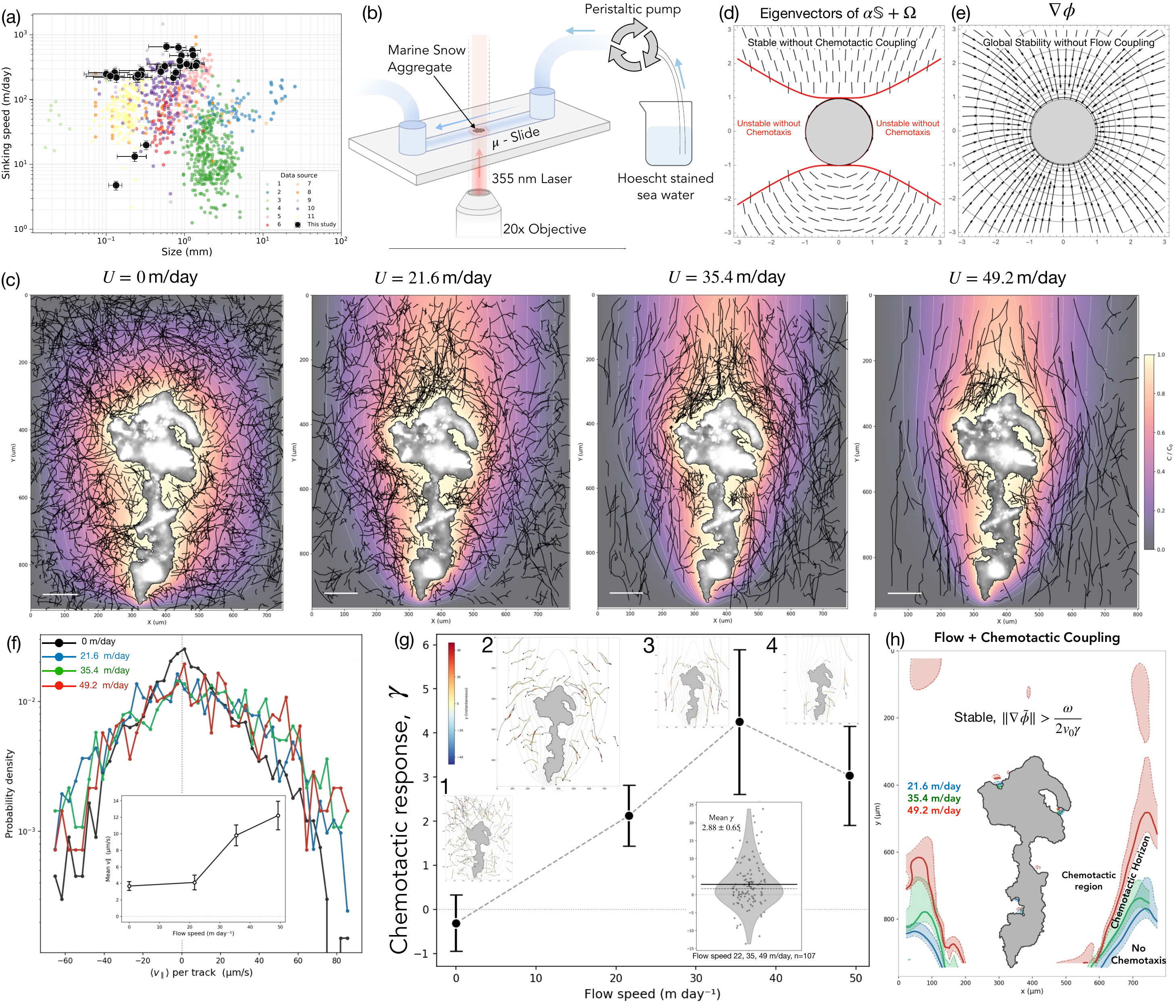}
 \caption{\textbf{Emergence of a microbial cloud and its chemotactic response:}(a) Marine snow sinking velocity as a function of particle size collected using vertical tracking microscope for the same sediment trap, compared with data compiled from past literature. No clear relationship exists in size and sinking speed \cite{chajwa2024hidden}. (b) In order to observe microbial clouds around marine snow particles, we set up a flow imaging system with flow velocity as a proxy for sinking speed. (c) Single cell trajectories of swimming microbes around marine snow with increasing flow $U=0, 21.6, 35.4, 49.2$m/day (left to right) with background chemical field acquired by solving the advection diffusion (AD) equation using diffusivity of oxygen [see methods D for details]. To disentangle the effect of flow-induced and chemotactic rotation, we perform stability analysis of microbial orientations (elongated shape, $\alpha=1$) for the inner solution of AD equation at low Pe and zero Re, with (d) just flow response  and (e) with just chemical response [see Supplementary]. (f) Probability density of the $\langle v_{||}\rangle$ as a measure of chemo-attraction towards marine snow [see Fig. 2(h) for the definition] for various flows, and the inset shows that the mean of the distribution increases with the flow speed. (g) Chemotactic response $\gamma$ (\ref{eqnX6}) measured for the top 20\% of the swimmers ranked according to the average value $\langle v_{||}\rangle$ measured along each track, shown in the inset as 1. 2. 3. 4 for increasing imposed flow $U=0, 21.6, 35.4, 49.2$m/day, respectively. The instantaneous $\gamma$ is measured using Equation \eqref{eqngamma}, for these highly chemotactic swimmers. Inset shows the distribution of $\gamma$ for $U\neq0$ with mean $ \langle \gamma \rangle  = 2.88 \pm 0.65$. (h) The competition between local chemotaxis (through $v_0 \gamma$) and vorticity $\omega$ yields a region where orientation can be stablized by chemotactic response, through the condition $\lVert \nabla \tilde{\phi}\rVert > \omega/2v_0 \langle \gamma \rangle$, showing the boundaries which separates regions of chemotaxis and no-chemotaxis for $U=21.6,35.4,49.2$m/day in blue, green and red color respectively, with $v_0$ is measured median speed across tracks, and variance in $\langle\gamma\rangle$ gives the shaded bands.
 }
 \label{Figure3}
 \end{center}
 \end{figure*}

 \section{Probing the Chemotactic Horizons with Flow}
 Before using imposed flow for probing the strength of chemotactic response, we turn to measuring its signature without flow. Although our assay consists of a natural populations of hundreds of distinct microbial species responding to a cocktail of compounds released by marine snow, for a baseline analysis we focus on a highly diffusive molecule released by marine snow \cite{Plough2001} that heterotrophic bacteria readily respond to \cite{Bell1972, blackburn1998microscale, Thar2001}. We solve the 2D diffusion equation for the oxygen concentration $\phi$, with fixed concentration $\phi_0$ at the aggregate boundary measured via thresholding of the binary image at t=0. For each track, we measure $\mathrm{v_{||}}$, the component of instantaneous speed along $\nabla \phi$ [see Fig. \ref{Figure2} (h)] and average it over the track time, to quantify the propensity of microbes to swim uphill the chemical gradient. We find a weak chemo-attraction in the absence of flow, with mean $\mathrm{v_{||}} = 2.72 \mu$m/s.

 We measure the sinking speeds of marine snow collected from the same sediment trap sample, using a vertical tracking microscope [see \cite{chajwa2024hidden} for the method]. Based on this and previously published data from \cite{chajwa2024hidden, Cael2021} [Fig. \ref{Figure3} (a)], we decide the range of imposed flows in our experiments 0-100 m/day [Fig. \ref{Figure3} (b)]. These flows mimic the sinking velocity of a marine snow particle. Our analysis pipeline allows us to track swimmers at single cell resolution for U = 0, 21.6, 35.4, 49.2 m/day. The non-motile population follow the streamlines, and used as PIV tracers to measure the flow using an iterative scheme that separates swimming and non-swimming population [see methods section C]. The resulting flow is used to calculate the oxygen profile by numerically solving a steady-state 2D scalar advection-diffusion equation \cite{leal2007advanced} with Dirichlet boundary condition on the aggregate boundary through a direct matrix inversion [for details see methods section D]. This problem has also been previously investigated for solute plumes of sinking spheres with non-porous \cite{Kiorboe2001FluidMotion} and porous \cite{Borer2022PorousMS} geometries.

With increasing flow speed we find that the swimmers accumulate downsteam of the aggregate forming a motility cloud for slow to moderate flow speeds. This active cloud sheds as the flow speed increases beyond a critical value [see Fig. \ref{Figure3}(c)]. The motile tracks are defined based on their departure from the streamline. We also quantify a threshold value of the velocity component perpendicular to the flow, determined using the noise floor of the PIV measurement [see Methods]. In our experiments, the swimmers respond passively to shear induced by flow and actively by chemotactic coupling. The stability analysis of the orientations at low Reynold and small P\'eclet numbers for spherical aggregates [see supplementary] with $\alpha=1$ (highly elongated swimmer) readily shows a vorticity dominated lateral region with no flow-induced alignment \cite{Joshi2025, Slomka2020}, while a phantom swimmer that responds only to chemical field and not flows would display alignment along $\nabla \phi$. This allows for the possibility of competition between vorticity and chemotaxis which together determine the net drift towards a marine snow particle [Fig. \ref{Figure3} (d) \& (e)].

To isolate and characterize this subtle chemotactic drift, we study in detail individual tracks of highly chemotactic microbes, top $20\%$ from the distribution of track-wise $\langle \mathrm{v}_{||}\rangle$ [Fig. \ref{Figure2} (h) \& Fig. \ref{Figure3}(f)]. These tracks [see supplementary Fig.6] qualitatively reveal nearly ballistic motion towards the aggregates with a near continuous steering through kicks of chemotactic torques, which motivates our model choice. Continuous steering in marine bacteria has been characterized in culture samples for Thiovulum majus \cite{Thar_2003}, but here is the first report of this behavior in open ocean samples and in the context of marine snow.  

The behavioral complexity of microorganisms in natural seawater samples [\ref{Figure2} (a)] presents a significant challenge in modeling the many diverse strategies organisms use to find and colonize sinking marine snow particles. We construct a low-dimensional model that captures the coupling between motility, flow, and chemotaxis and study how these factors make it easier or harder for a microbe to find a sinking marine snow particle. To unambiguously separate active swimming behaviors from passive hydrodynamic effects, we turn to the equations of motion of an active brownian particle (ABP) \cite{romanczuk2012active} with swimmer position $\mathbf{X}$ and orientation $\mathbf{p}$.  An intrinsic swimming velocity $\mathrm{v}_{0}$ and background flow velocity $\vect{U}$ enters the position dynamics, $\dot{\mathbf{X}} = \vect{U} \,+ v_{0} \mathbf{p}$. We endow the orientation equation with a chemotactic torque term $v_{0} \gamma \mathbf{p} \times (\nabla \bar{\phi} \times \mathbf{p})$, coupling $\mathbf{p}$ to $\nabla \phi$ in addition to the flow gradients through the symmetric and antisymmetric parts, denoted by $\bsf{S}$ and $\vect{\Omega}$, respectively \cite{Pedley1992}, 

\begin{equation}
\dot{\mathbf{p}} = [\mathbb{I} - \vect{p}\vect{p}]\cdot(\alpha \bsf{S} + \vect{\Omega})\cdot \mathbf{p} \, + \, v_{0} \gamma \mathbf{p} \times (\nabla \bar{\phi} \times \mathbf{p}) \, + \, \text{noise},
\label{eqnX6}
\end{equation}
where the projection operator $\mathbb{I} - \vect{p}\vect{p}$ in \eqref{eqnX6} ensures unit magnitude of $\vect{p}$. The chemotactic coupling in \eqref{eqnX6} is permissible on general symmetry ground \cite{Suropriya2014}, and breaks the fore-aft symmetry of the swimmer $(\vect{p} \to -\vect{p})$. Note that in \eqref{eqnX6} we neglect polar flow-alignment \cite{Maitra2014}, and the confinement-induced breakdown of Galilean invariance \cite{Brotto2013}.  We focus on highly chemotactic microbes to measure our chemotactic turning response $\gamma$ which differs from the chemotactic sensitivity $\chi$ commonly used in Keller–Segel-type descriptions of bacterial chemotaxis \cite{KellerSegel1971}.

Below, we first directly measure the chemotactic response $\gamma$ of individual microbes to a chemical gradient field of the aggregate using the deterministic sector of the full dynamics, where the noise manifests in the measured distribution of $\gamma$. Second, we use the measured $\gamma$ to demarcate the region around the aggregate in which chemotaxis overturns vorticity induced rotation. 

The time derivative of position equation along with \eqref{eqnX6} gives the dynamical equation for swimmer's velocity in a steady-state flow [see supplementary section III], 
\begin{align}
    \frac{d\vect{v}}{dt} &=  \vect{v}\cdot \vect{\nabla U} + \vect{\Omega} \cdot (\vect{v} - \vect{U}) + \gamma v_{0} ^{2} \vect{\nabla} \bar{\phi} \nonumber \\ 
    & - \gamma \left(\vect{v} - \vect{U}\right) \left(\vect{v} - \vect{U}\right) \cdot \vect{\nabla} \bar{\phi} \, + \, \text{noise}.
\end{align}
Taking the inner product of both sides with $\nabla \phi$, and noting that $(\vect{v} \cdot \nabla) \vect{U} = \bsf{S} \cdot \vect{v} + \vect{\Omega}^{\top} \cdot \vect{v}$, and the skew-symmetry of $\vect{\Omega}$, gives the non-dimensional chemotactic coupling,
\begin{equation}
    \gamma = \frac{(\dot{\vect{v}} \, - \,  \bsf{S} \cdot \vect{v} \, + \vect{\Omega} \cdot \vect{U} ) \cdot \nabla \bar{\phi}}{v_{0}^{2} \lVert \nabla \bar{\phi} \rVert^{2} \sin^{2} \delta } \, + \, \text{noise},
    \label{eqngamma}
\end{equation}
 where $\delta$ is the angle between swimmer orientation and $\nabla \phi$, rendering $\gamma$ ill-defined in the limit $\delta  \to 0$. Note that the Lamb vector $\vect{\Omega} \cdot \vect{U}$ is $\mathcal{O}(U^2)$ and thus small for slow viscous flows. Using \eqref{eqngamma} we measure the chemotactic response from the acceleration $\dot{\vect{v}}$ and velocity $\vect{v}$ of the tracks and the local flow velocity $\vect{U}$, shear $\bsf{S}$ and the numerically calculated chemical gradient $\nabla \bar{\phi}$, in a time-averaged sense, $\langle \gamma \rangle \equiv 1/T\int_{0}^{T} \gamma dt$ along a trajectory. This quantity is ill-defined in the limit $\delta \to 0$, that is, when the swimmer orientation aligns with $\nabla \phi$ [see method section E, Fig. \ref{FigureS3}(e)-(h)]. The measured $\gamma$ [Fig. \ref{Figure3}(g)] should be seen as being close to the upper bound on chemotactic response.

In the zero-noise limit, Equation \eqref{eqnX6} can be recast as $\dot{\vect{p}} = (\mathbb{I} - \vect{pp})\cdot [ (\alpha \bsf{S} + \vect{\Omega}) \cdot\vect{p} \, + \, v_{0}\gamma \, \vect{\nabla} \bar{\phi} ] $, which yields the steady-state solution for swimmer orientation [see Supplementary for details],
\begin{equation}
\vect{p} =  v_{0} \gamma [\lambda \mathbb{I} - (\alpha\bsf{S} + \vect{\Omega}) ]^{-1} \cdot \vect{\nabla} \bar{\phi}.
\label{eqn:steadyp}
\end{equation}
To match the information content of our field collected swimmer data, which lacks direct shape measurements of cells, we set $\alpha =0$ (spherical limit of prolate spheroid). For spherical swimmers the stable region in Fig. \ref{Figure3} (d) collapses on a vertical line passing through the aggregate center, producing globally ($\mathbb{R}^2 \setminus \{x=0\}$) unstable orientations. We therefore study the combined response to flow and chemical field where the chemotactic alignment competes with local vorticity $\omega = 3U x/2ar^{3}$. Imposing $\lVert \vect{p} \rVert = 1$ identifies $\lambda$ and yields a condition for stable orientations $\lVert \nabla \bar{\phi} \rVert \geq \omega/2v_{0} \gamma$. The resemblance of the equality with the Eikonal equation of geometric optics is superficial, since $\phi$ is determined independently via the advection-diffusion equation. The dominant term in both $\lVert \nabla \phi \rVert$ and $\omega$ is $\mathcal{O}(1/r^{2})$, presenting competing effects. The region where the slope of the chemical profile is greater than $\omega/2v_{0} \gamma$ has locally stable orientations, wherein microbes can efficiently display chemotaxis towards the aggregate [see Fig. \ref{Figure3}(h)]. The chemotactic horizon does not exist around a non-sinking particle; it emerges with flow.
\begin{figure}[t]
 \begin{center}
 \includegraphics[width=8.5 cm]{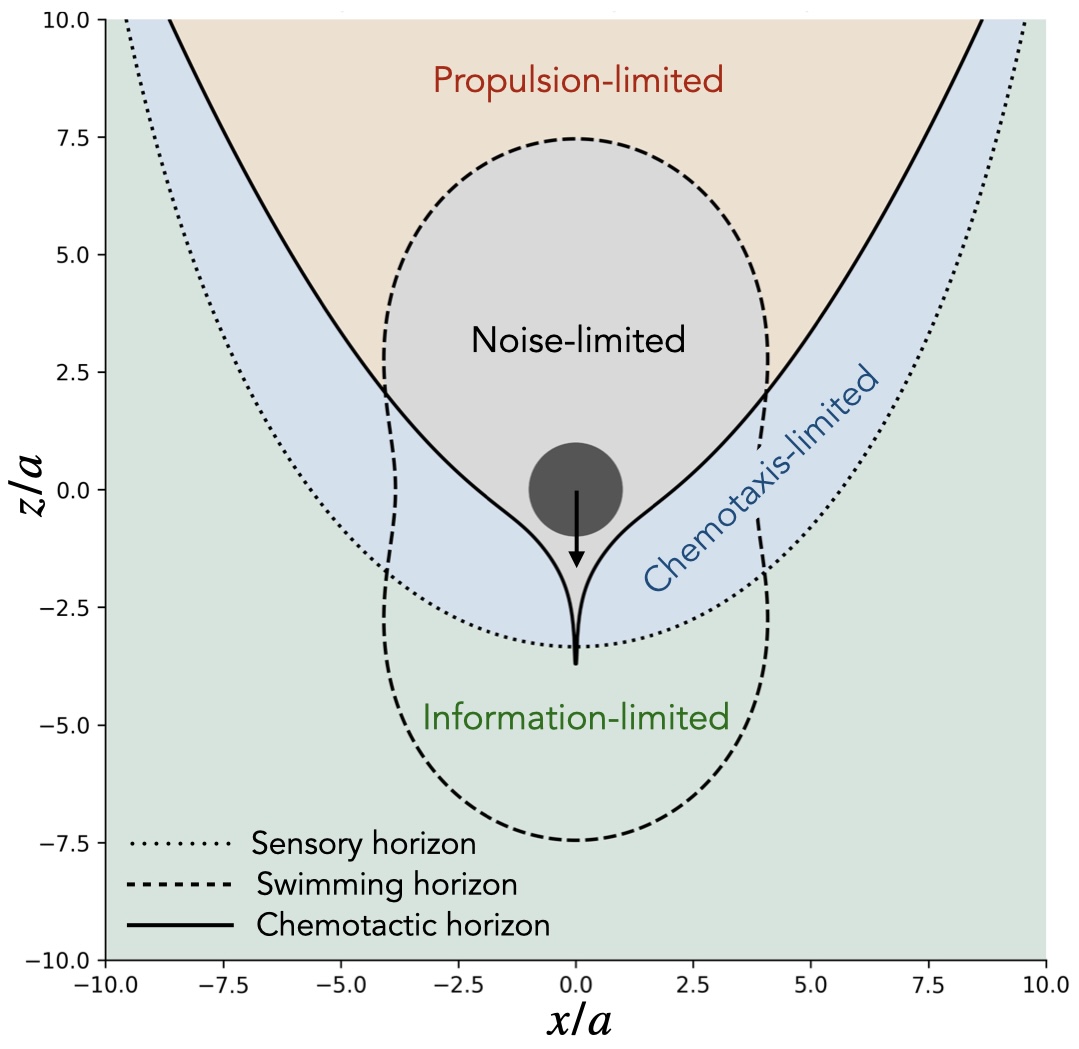}
 \caption{ \textbf{Dynamical phase diagram of microbial encounter with a sinking particle}: zero Reynolds, low Peclet limit of the flow and chemical profile, for typical paramerers [see supplementary] gives rise to: (1) sensory horizon as described in \cite{Berg1977}, (2) chemotactic horizon given by the balance of chemotaxis and vorticity with measured chemotactic turning response ($\gamma = 2.88$), and (3) swimming horizon from the balance of vertical component of Stokes settling-induced flow and swimming speed. These boundaries gives rise to various zones representing mechanisms that limit encounter between microbe and sinking particle.}
 \label{Phase_diagram}
 \end{center}
 \end{figure}

Spatial extent of the chemotactic horizon defined above is determined by a non-dimensional group $\Gamma = \gamma \mathrm{v}_{0}/b\omega$, where $b$ is the size of the swimmer. 
Being within the chemotactic horizon is a necessary but not sufficient condition for encounter, akin to the classic sensory horizon \cite{Berg1977} with the non-dimensional group $v_0 T^{3/2}(4 \pi D \phi_0/b)^{1/2} \equiv \kappa$, where $T,D,\phi_0$ are the signal integration time, diffusivity and chemo-attractant concentration at the aggregate surface, respectively. Furthermore, to successfully chase a sinking particle it is necessary for the microbe to bypass a critical swimming to sinking speed ratio, which is another non-dimensional group $v_0/U \equiv \beta$.

To conceptually understand the relative extent of these effective microbial transport boundaries that fundamentally govern encounters, we analytically calculate these limiting curves at low P\'eclet and zero Reynolds number, in the outer region of the Acrivos-Taylor solution \cite{leal2007advanced} of the advection-diffusion equation, to leading orders in $Pe$. This analysis reveals a dynamical phase diagram comprising distinct zones, each corresponding to a mechanism that limits encounters between a microbe and a sinking particle. These mechanisms are [see Fig. \ref{Phase_diagram}]: (1) sensing limited \cite{Berg1977}, described by $\kappa \lVert \nabla \phi\rVert/\sqrt{\phi} \geq 1$, (2) chemotaxis limited, $2v_0 \gamma \lVert \nabla \phi \lVert/\omega \geq 1$, and (3) propulsion-limited, where the swimmer lacks sufficient speed to catch the particle, as determined by the level curve of the vertical component of the Stokes flow around a sinking sphere, $ u_z \leq \beta$. Here, $\nabla, \phi$, and $u_z$ are non-dimensionalized using $b$, $\phi_0$ and $U$, respectively. In the grey region, where all three conditions are satisfied, the encounters are limited only by the inherent stochasticity of microbial dynamics. This phase-diagram also provides a qualitative explanation for the existence of a critical sinking speed $v_c$ beyond which the grey region in Fig. \ref{Phase_diagram} collapses, thereby suppressing encounters.
 
 \section{Microbially Regulated Threshold Dynamics of Marine Snow Transport}

\begin{figure*}[t]
 \begin{center}
 \includegraphics[width=18.0 cm]{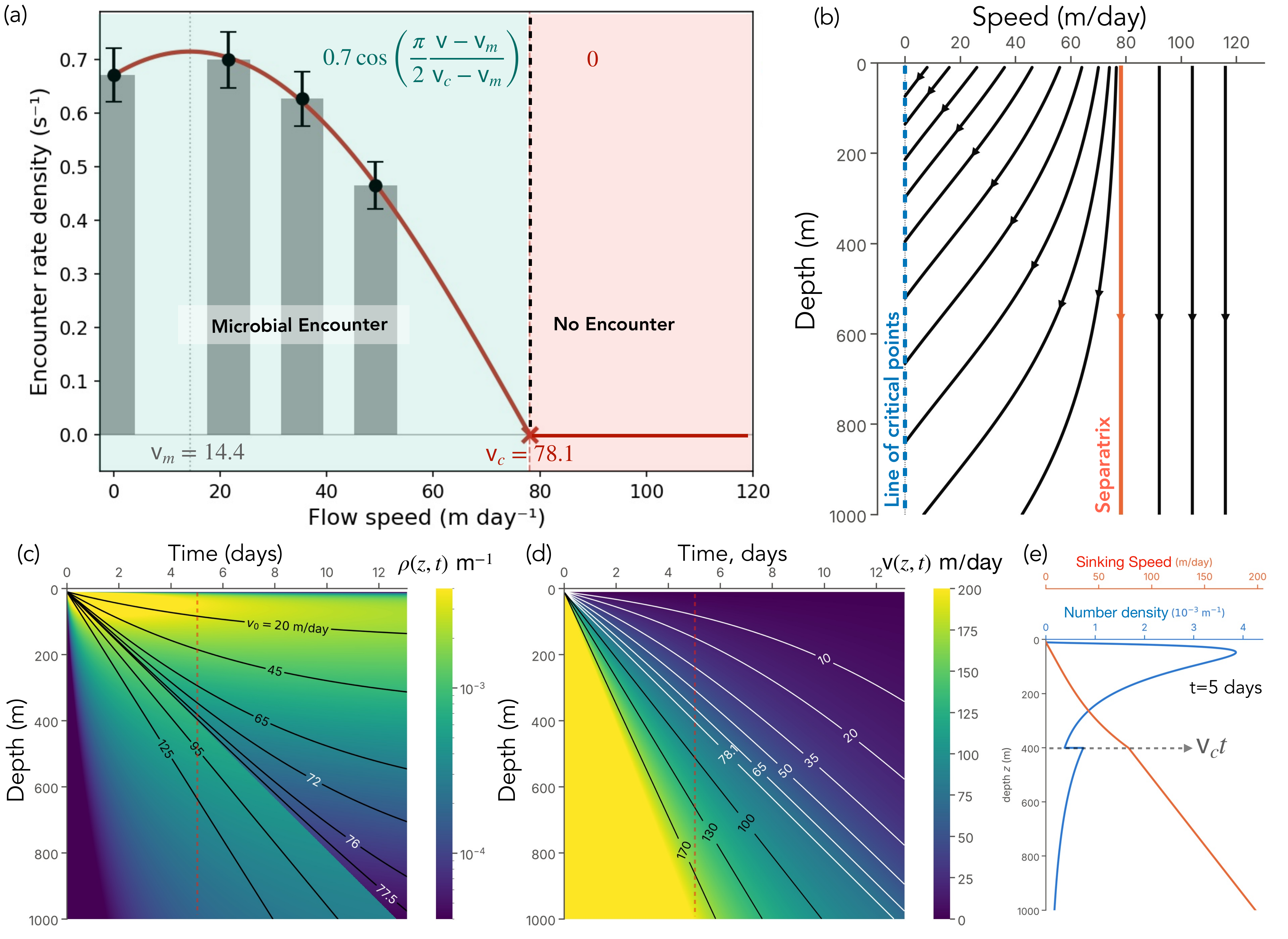}
 \caption{ \textbf{Threshold dynamics of sinking marine snow:} (a) Encounter rate density defined as the number-density of swimmers within the $0.9\phi_0\leq \phi \leq \phi_0$ normalized by the background swimmer number-density, as a function of flow speeds. A cosine fit gives the speed $v_m = 14.4$m/day at which the encounter peaks and a threshold speed $v_c = 78.1$m/day at which encounter diminishes to zero. (b) Assuming bacterial encounter and colonization as the primary process for marine snow degradation [Supplementary Fig. S4], we construct a phase portrait in $(v,z)$ plane showing the characteristic curves derived analytically from the Stokesian settling dynamics with velocity dependent degradation rate [Eqn. \eqref{thresholdEqn}]. The blue dotted line shows the line of critical points marking the asympotically halting particles at a depth $z_\text{max}$ which diverges as $v_0 \to v_c$ marking the separatrix (red line) splitting the phase-space into over-damped and ballistic transport regimes. (c) Number-density field $\rho(z,t)$ calculation from the solution of the kinetic transport equation for an initial pulse of particles initialized at depth $z_0$ with a given lognormal distribution of initial speeds $v_0$: $f(z_0,v_0) = N \delta(z-z_0)\mathit{LN}(v_0,\mu, \sigma^{2})$ with characteristics shown for various $v_0$, and (d) the velocity field comes from solution of the characteristic equation. (e) Number-density field (blue) and settling speed (red) at t=5 day transect (red dotted line in (c) \& (d)) plotted as a function of depth emulates the fate of marine snow flux event when remineralization is primarily shaped by microbial encounter based colonization and subsequent degradation.}
 \label{Figure4}
 \end{center}
 \end{figure*}
We quantify encounter rate as the time-averaged number density of microbes within the concentration band $0.9c_0 \leq c \leq c_0$, over an observation window ($T=30 \mathrm{s}$), normalized by the unperturbed ($U=0$) number density in the field of view, excluding the projected area of the aggregate. This normalized measure is equally valid in three dimensions, as the volume-thickness factor cancels in the ratio. Microbial encounter rates vary non-monotonically with effective sinking speed [see Fig \ref{Figure4}(a)]: they are enhanced at weak flows, peak at $v_m = 14.4$ m/day, and decline sharply as microbial dynamics become tracer-like, with negligible encounters beyond the critical speed $v_c = 78.1$ m/day. The initial enhancement is qualitatively consistent with Sherwood-number-based predictions for flow-enhanced diffusive encounters \cite{Kiorboe2002Colonization, Slomka2023EncounterRates}, which do not capture the observed non-monotonic dependence on speed. Although the sinking speed associated with peak degradation in \cite{Alcolombri2021} is close to $v_m$, the underlying mechanism differs fundamentally from the activity–flow coupling studied here.

 Consider a dilute sedimenting suspension of non-interacting marine snow particles drifting under gravity without other background flows. To illustrate the effect of velocity dependent degradation of particles due to optimal colonization by swimming bacteria, we assume an absence of fracture, merger, formation, and removal of particles, thus producing a number-conserving dynamics. We do not imply that these higher-order processes are not important, but here we are interested in understanding the fate of marine snow when microbial encounters are the dominant mechanisms for colonization and degradation. Through this simplification, we seek to construct a mechanistic model that connects nonlinearities present in remineralization at microscopic scale to large-scale transport. With this preamble, we construct the dynamical equations for a slowly drifting and degrading/reacting suspension of marine snow particles initialized in the upper ocean. The position-velocity $(z,v)$ phase space dynamics in terms of a single particle distribution $f(z,v,t)$ is governed by a 1D kinetic transport equation \cite{huang1987statistical},
\begin{equation}
    \frac{\partial f (z,v)}{\partial t} + \frac{\partial}{\partial z}[vf] + \frac{\partial}{\partial v} [- R(v) v f] =0,
    \label{Transporteqn1}
\end{equation}
where Stokesian sedimentation of individual particles evolves through first-order equations, $\dot{z} = v$ and $\dot{v} = -R(v)v $, which incorporates microbial degradation through a velocity dependent rate,
\begin{equation}
  R(v) = R_0 \times \begin{cases}
             \cos (qv-\theta_0) \,\, , \,\,0\leq v < v_c\\
              0 \, \, , \,\,v \geq v_c
            \end{cases}.
            \label{thresholdEqn}
\end{equation}
Here $q \equiv \pi/2(v_c - v_m)$ and $\theta_0 \equiv \pi v_m/2(v_c - v_m)$ are empirically determined based on the speed $v_m$ at which the microbe-marine snow encounter is maximum and a critical sinking velocity $v_c$ at which the encounter rate diminishes to zero. Our model assumes that microbial degradation dissolves the particulate organic matter (POM) in marine snow, replacing it with another neutrally buoyant fluid phase while leaving effective hydrodynamic mobility unchanged. Equivalently, volume fraction of POM $\phi$ evolves as $\dot{\phi} = - R(v) \phi$. We adopt a quasi-steady closure in which effective microbial degradative activity responds instantaneously to the encounter rate at a given sinking speed, neglecting memory associated with microbial colonization. A similar degradation rule, but without threshold, has been invoked previously \cite{Omand2020SinkingFO} to understand Martin curve. The measured encounter-rate curve prescribes the velocity dependence of R(v), while its amplitude is set to $R_0=0.13,\mathrm{day}^{-1}$, using the mean carbon-specific respiration rate measured for laboratory-produced aggregates \cite{IversenPloug2010}. We use this respiration rate as a proxy for the fractional POM degradation rate.

Note that remineralization enters as a non-conservative force on particles, which renders the phase-space volume compressible in the subcritical region ($v_{0}<v_c$). The phase portrait can be solved analytically for the subcritical and the supercritical ($v_0 \geq v_c$) regions separately. Subcritical dynamics yield a maximum depth $z_{\text{max}} = z_0 + \ln|(\sec (qv_0 - \theta_0) + \tan(qv_0 - \theta_0))/(\sec \theta_0 - \tan \theta_0)|$,
forming a line of critical points [blue line in Fig.4(b)]. At criticality, as $v_0 \to v_c$ from the left, $z_{\text{max}}$ diverges as $\log \left| (1 - v_{0}/v_c)^{-1} \right|$, marking the separatix [red line in Fig. \ref{Figure4}(b)]. Supercritical dynamics show undamped speeds, thanks to the lack of microbial encounters and subsequent degradation, giving rise to vertical rays in the phase portrait.

Departing from the steady-state paradigm in marine snow transport \cite{Giering2017} which does not match the fundamentally unsteady nature of primary production in the ocean, we study how an initial pulse of marine snow particles produced at $z_{0} = 10m$ with a log-normal distribution of initial speeds, $f(z,v, t=0) = N \delta(z-z_{0}) \text{LN}(v, \mu, \sigma^2)$ evolves in time via \eqref{Transporteqn1} and \eqref{thresholdEqn}. The transport equation \eqref{Transporteqn1} is solved semi-analytically for number-density field, $\rho(z,v,t) = \int_{0}^{\infty} f(z,v,t) dv$ and velocity field $v(z,t) = \int_{0}^{\infty} f(z,v,t)dz$ using the method of characteristics [see Supplementary]. $\rho(z,t)$ distinctly shows a jump discontinuity at $z= v_c t$ and $v(z,t)$ shows the splitting of the particle population into slow and fast sinking cohorts in the upper and deep ocean, respectively. While a sharp jump-discontinuity in $\rho(z,t)$ is expected to be smoothened in natural systems, we expect the bimodality of the number density field and the spatial splitting of fast ($v_{0}>v_c$) and slow ($v_{0}<v_c$) sinking populations to be robust and observable consequences of the threshold dynamics. This fundamentally changes how we interpret time-averaged sediment-trap collections as a measure of time-dependent vertical carbon flux: particles collected at a given depth and time may have experienced very different histories and may therefore have very different eventual fates. 

\section{Conclusions}

Current models of the biological carbon pump present gigaton scale uncertainty in our climate models \cite{Cael2024TinyCU}, which can be largely attributed to the lack of predictability of marine snow transport from the surface to the deep ocean as it sinks through a sea of microbes that colonize and degrade it. This produces an attenuation of flux $F$ with depth, modeled as power a law decay $F=F_0(z/z_0)^{-b}$ (Martin curve) \cite{martin1987vertex} where all the micro-scale mechanisms (marine snow sedimentation and microbial encounter and degradation) are lumped into a constant exponent $\beta$. Sedimentation with remineralization is a convolved process that involves internal dynamics, such as fluid-structure interaction in marine snow \cite{chajwa2024hidden}, local chemistry and scalar diffusion \cite{Plough2001, Kiorboe2001FluidMotion}, and microbial composition with perpetual interactions between microbes and marine snow \cite{Kiorboe2002Colonization, Stocker2008}. These processes together shape the exponent $\beta$ \cite{Cael2021}, assuming there exists a power-law scaling with a constant exponent. Focusing on encounters, we find that how microbes find marine snow (first step of degradation) is a convolved mechanism involving several internal biological knobs through which microbes sense and navigate the heterogeneous fluid mechanical and chemical oceanscape. Our work provides a direct observational window into this enigmatic process.

First, we probe this black-box by directly observing the true dynamical complexity of microbe-marine snow interactions, as it is, via direct live imaging and tracking of microbes around freshly collected marine snow in controlled flows at single-cell resolution on a research vessel in the open ocean. Beyond tackling the challenge of imaging single bacteria in an open ocean setting on a research vessel, we also develop a custom analysis pipeline that combines algorithmic tracking with human-assisted frame-by-frame error correction of each track. Although extremely tedious, human verified tracks reveal the rich individual microbial dynamics and statistics that can be trusted. The resulting tracks reveal that, 1) nearly half of the microbes were not swimming at any given time, and 2) a diverse microbial motility is present, with a few distinct classes that visibly stand out, including (a) ballistic dynamics, (b) burst-pause, (c) run-reverse-flick, (d) run-stop-run, (e) hovering, and (f) diffusive [see Fig. \ref{Figure2}(a)]. The burst-like motility is the most remarkable aspect among these diverse traits as it marks the transitions between motile and non-motile subpopulations, rendering it a non-equilibrium steady state [supplementary video 3]. How the statistics of this ensemble depends on change in environmental cues or internal states remains to be investigated. Furthermore, we measure size distribution of these diverse microbes using the fluorescent signal as an indirect proxy, which allows us to provide a rough estimate of the energetic cost of swimming, as a metabolic fraction, for natural microbial assemblage in the ocean. Simultaneous measurement of the microbial size, orientation, shape and flagella are needed to refine this energy estimate and enable quantification of shape–hydrodynamic coupling \cite{Jeffery1922} within our framework. The individual microbial tracks are limited by their transit time through the focal region, providing only a two-dimensional projection of the full dynamics, an aspect that calls for 3D high-speed imaging techniques such as phase-contrast \cite{taute2015high} and holographic \cite{Kim2024} methods. Field-deployable stable microscopy platforms will improve our measurement precision.

Second, using controlled flows corresponding to ecologically relevant marine snow sinking speeds, we extensively study the turning response of highly chemotactic swimmers to chemical gradient field generated by a marine snow particle, inferred numerically from measured flow fields. We introduce a chemotactic response parameter $\gamma$ through a symmetry-based phenomenological model of micro-swimmers \cite{Suropriya2014} which incorporates swimmer orientation as an internal dynamical variable that responds to fluid shear and the chemical field. Here, we use oxygen as a chemo-attractant; however, natural marine snow emanate a truly multi-component scalar fields with range of moelcular diffusivities, to which microbes can respond differentially \cite{Bloxham2026}. This necessitates direct simultaneous measurement of chemical field through spectroscopy methods, such as \cite{Kamp2024}. Our model separates deterministic dynamics from stochastic fluctuations, while recognizing that both directed motion and active fluctuations arise from the same underlying processes: flagellar elasto-hydrodynamics \cite{Son2013} and biological signal processing \cite{Berg1977}. These processes can generate adaptive swimming speeds and turning responses through memory-kernel-based information processing, which remains largely unexplored beyond E. coli.

Third, by directly measuring encounter-rates of the ensemble of microbes with marine snow in controlled flow speeds $U$, we find that encounter rate vary non-monotonically with flow speeds; increasing with flow, peaking at $v_{m}=14.4$ m/day before sharply declining to near-zero for $U\geq78.1$ m/day. The speed corresponding to maximum encounter rate is close to that from \cite{Alcolombri2021} which incorporates erosion; and it remains to be explained how encounter and erosion together shape degradation and resulting fluid-structure interactions \cite{Shelley2012} in the marine snow `comets'. Furthermore, how the biological viscoelastic phase in marine snow \cite{chajwa2024hidden} alters the microbial encounter dynamics remains unexplored. We incorporate the non-monotonic critical speed-dependent encounter, with an encounter-limited remineralization hypothesis, into a kinetic transport model which we solve semi-analytically for a pulse of marine snow with log-normally distributed speed. We find a sepratrix in the phase-space dynamics of marine snow demarcates overdamped and ballistic trajectories, with velocity segregation -- slow sinking particles populate the upper-ocean, being retained there, while fast sinking particles fall to the sea-floor without appreciable remineralization. In this treatment we assume a quasi-static closure in which degradation is directly proportional to encounter-rate [see Supplementary]. Future studies are needed to rigorously coarse-grain microbial ecology \cite{Crocker2025} on sinking particles to precisely map encounter to degradation, with steady in-flux and out-flow of microbes through the activity-flow-chemotaxis coupling discussed here. Our work highlights the intricate coupling between microhydrodynamics and microbial ecology in marine snow transport, motivating further investigation of their interplay. 

Lastly, how the changing climate is perturbing marine microbial consortia \cite{peixoto2026global}, and in-turn the ocean-carbon sequestration, remains a pressing open problem, and this work is a small step towards real-time ocean microbial diagnostics.











\section{Methods}
\subsection{Tracking individual microbe}
Single cell tracking for data collected in open ocean conditions is challenging. Thus, significant effort was given to manually curate the data and provide the best single cell tracking raw datasets for analysis. This analysis involved a combination of algorithmic tracking and human curation to verify every single track used in this work. In total $\simeq 10^{4}$ tracks ($\geq 4$ spots) were isolated for single cells across flow scenarios, which provided us a broad range of motility motifs. No assumption was made regarding the phenotype or behavior of single cells and an unbiased tracking approach was best suited for field collected data for heterogeneous samples.  Briefly, the raw images were imported to FIJI and their contrast was enhanced for fluorescent signal from the cells as compared to the noise floor. Raw tracks were produced in Fiji's TrackMate, with spot detection using Laplacian of Gaussian (LoG) and linking using Advanced Kalman Tracker \cite{Jaqaman2008RobustSP} with optimized parameters. The resulting tracks were subjected to a multi-stage quality-control pipeline before analysis which included the following steps. First, any spot whose pixel coordinates fell within the filled binary mask of the aggregate interior was removed from the track so as to not confuse microbes that had already encountered the marine snow as compared to first encounter (our focus). The remaining spots were reassembled into sub-tracks, splitting automatically wherever the boundary-filtering step had introduced a frame gap greater than one frame. Sub-tracks shorter than four spots were discarded. The subsequent cleaning steps differed between no-flow and flow experiments, outlined below: 

\textit{For no-flow experiments}, 
The tracks were ranked for review by a jump-detection algorithm. For each track the per-step displacement was computed, and steps were flagged as jumps if they exceeded both an absolute threshold ($10 \mu$m) and a ratio threshold ($5 \times$ the track’s median step displacement). Tracks containing at least one such jump were designated `suspicious' and sorted to the front of the review queue in descending order of maximum jump ratio; non-suspicious tracks followed, ordered by decreasing length.

The GUI displayed each track in two panels: a cropped fluorescence image (padded 150 px around the track bounding box, with the trajectory overlaid in cyan) and a bar chart of per-step speeds with the jump-detection threshold marked. The reviewer could scrub through frames to watch the particle move. For each track, one of five decisions was recorded: Keep (accept as-it-is), Discard (remove entirely, for tracks that had random jumps between cells), Split (divide into two sub-tracks at the current frame or at automatically detected jump frames), Split and discard prefix (keep the tail segment only), or Split and discard suffix (keep the head segment only). Split sub-tracks were retained only if they contained $\geq4$ localizations. Decisions were saved incrementally to a JSON file, allowing the session to be resumed.

Tracks could also be tagged with qualitative labels (e.g., slow $\to$ fast, reversal, artifact, hovering) or free-text notes for downstream stratified analysis. Tags were propagated automatically to sub-tracks produced by splitting.

\textit{For flow experiments}, a PIV-derived flow field was used to pre-screen each track automatically before any manual step. Tracks were classified as passive streamers if their mean cosine alignment with the local flow vector was $\cos \theta \geq 0.8$ and their root-mean-square perpendicular speed was $\langle \mathrm{v}_{\perp} \rangle_{rms}\leq 24.3 \mathrm{\mu m s^{-1}}$, a threshold set as the $95^{\text{th}}$ percentile of the $\mathrm{v}_{\perp}$ noise floor distribution of tightly flow-aligned tracks ($\cos\theta \geq 0.9$). Tracks were classified as erratic if their maximum-to-median step-size ratio exceeded 4 or their mean turning angle exceeded $\pi/2$. Tracks that satisfied neither criterion were presented in the flow-aware GUI for the same keep/discard/split manual review described above. The automatic classifications (passive or erratic) were recorded as tags alongside each manually reviewed decision.

\subsection{Swimmer size estimation}
The purpose of this estimate is to provide a rough scale for cell size. Cell sizes were estimated by fitting a 2D elliptical Gaussian to the fluorescence intensity profile of each tracked bacterium at its best-focus frame. Since DNA in bacterial cells (nucleoid) approximately fill the whole cell volume \cite{Gray2019}, utilizing fluroescent imaging of DNA is a good approximation for cell size/volume. For each track, the spot with the highest TrackMate quality score (a proxy for focus sharpness) was selected from among the top quartile of quality values along that track. A $33 \times 33$ pixel cutout ($\pm16$ px) was extracted from the corresponding raw image frame, centered on the TrackMate-measured position. The intensity profile was modeled as
$I(x,y) = A \exp [-(x_r/\sigma_x)^2/2 - (y_r/\sigma_y)^2/2] + B$, where ($x_r, y_r$) are coordinates rotated by angle $\theta$, and A, ($x_0, y_0$), $\sigma_x, \sigma_y, \theta, B$ were all free parameters. The centroid was constrained to $\pm2$ pixels from the TrackMate position to allow sub-pixel refinement while preventing the fit from jumping to a neighboring cell. Fits were accepted only if the fitted amplitude exceeded 3 counts and the relative uncertainty on both $\sigma$ values (from the covariance matrix) was below 50\%.
The minor axis $\sigma_\text{minor} = \text{min}(\sigma_x, \sigma_y)$ was used as the cell-size proxy because it is least affected by motion blur (which elongates the point spread function (PSF) along the direction of motion and inflates $\sigma_\text{major}$). The apparent minor-axis width was corrected for the diffraction-limited PSF in quadrature: $\sigma_\text{cell} = \sqrt{(\sigma_\text{minor}^2 - \sigma_\text{PSF}^2)}$ with $\sigma_\text{PSF} = \lambda/(2 NA) = 0.46 \mu m / (2 \times 0.5) = 0.46 \mu m$, using $\lambda = 0.46 \mu m$ (Hoechst emission peak) and known numerical apperture NA = 0.5. Tracks for which $\sigma_\text{minor} < \sigma_\text{PSF}$ (physically implying the cell is below the resolution limit) were recorded as NaN and excluded from size statistics. Two additional quality filters are applied on signal to noise ratio ($SNR >3.0$) and to exclude out of focus ``blobs" from spherical aberration ($\sigma_{\text{min}}>2.2 \mu m$) [see Supplementary size data for sizes].

\subsection{Measuring flow field: Non-motile fraction as a PIV probe}
The background flow field in each flow condition was extracted from the raw particle tracks using an iterative ensemble-PIV approach. Track positions exported by TrackMate are already in $\mathrm{\mu m}$; per-frame displacements $(\Delta x, \Delta y$ in $\mathrm{\mu m})$ were computed between consecutive frames for each track segment (only for contiguous frame pairs, i.e. frame gap = 1).

A regular spatial grid was defined with a window of 120 pixels ($\sim40 \mu m$) and a stride of 40 pixels ($\sim13 \mu m$). At each grid node the mean displacement of all track steps whose start position fell within the rectangular window was taken as the local flow velocity, provided at least 5 steps contributed (otherwise the node was left undefined). Undefined nodes were filled by iterative nearest-neighbour Gaussian inpainting (15 iterations, $\sigma = 1.5$ px) followed by a final Gaussian smoothing pass ($\sigma = 2.5$ px). The resulting 2D velocity field was bilinearly interpolated \cite{chajwa_activeparticles_2026} at arbitrary positions for downstream use.

The procedure was iterated (3 passes): the first pass used all tracks to compute a preliminary flow field; subsequent passes used only the passive-streamer subset ($\langle \mathrm{v}_{\perp}\rangle_{\text{rms}} \leq 24.3 \mathrm{\mu m s^{-1}}$) to progressively refine the flow estimate by removing the contribution of actively swimming bacteria. Iteration continued until the passive set converged. The PIV grid axes are in native image coordinates (row 0 = top, y increasing downward). Before being passed to the advection-diffusion solver, velocities were transformed to the solver’s convention (y increasing upward) by reversing the row order of the grid and negating the y-component of velocity.

\subsection{Numerically solving the advection-diffusion equation}
The chemical concentration field (normalized, $\phi/\phi_0$) around the aggregate was obtained by solving the steady advection-diffusion equation $D \nabla^{2}\phi - \vect{U} \cdot\nabla \phi = 0$,
on a 2D Cartesian grid derived from the segmented aggregate mask. The mask was downsampled so that its longest dimension spanned 260 grid cells, giving an effective pixel size = (native pixel to $\mu$m) $\times$ (long dimension/260). The aggregate was embedded in a domain padded by a factor of 1.0 (i.e. the domain extended by $\sim 50\%$ beyond the mask on each side to avoid boundary artifacts), giving a typical grid of $\sim 400 \times 400$ cells.
 
 \textit{Discretization}: Diffusion was discretized by the standard 5-point central-difference Laplacian (second-order accurate, $\mathcal{O}(\Delta x^2)$). Advection was discretized by a first-order upwind scheme, which is unconditionally stable for any P\'eclet number but introduces numerical diffusion of $\mathcal{O}(\Delta x)$. With D $= 2000 \mathrm{\mu m^2/s}$, typical flow speeds of $50-200 \mathrm{\mu m/s}$, and grid cell $\sim 1-2 \mu m$, the cell P\'eclet number $Pe_{\mathrm{cell}} = U \Delta x/D = 0.05-0.2$, so the scheme operates in the diffusion-dominated regime and numerical diffusion is negligible relative to physical diffusion.
 
 \textit{Boundary Conditions}: The aggregate surface was assigned Dirichlet $\phi = \phi_0 = 1$ (aggregate as chemical source). Non-outlet domain edges (upstream and lateral) were assigned Dirichlet $\phi = 0$ (ambient background). The downstream (outlet) edge was assigned a zero-gradient Neumann condition, allowing the chemical plume to exit the domain freely rather than being artificially forced back to zero at the edge of the imaging window. For the no-flow condition, where there is no preferred downstream direction, all four edges were set to Dirichlet $\phi =0$.

\textit{Solution Method}: The linear system (sparse matrix of size $\mathrm{N \times N}; \mathrm{N = N_y \times N_x}$) was assembled in CSR format and solved directly using a sparse LU factorization using python function scipy.sparse.linalg.spsolve. This gives the exact solution of the discretized system with no iterative residual. The numerical error is $\mathcal{O}(\Delta x)$ truncation error of the upwind advection scheme.
For flow conditions the velocity field u = (u, v) was supplied from the PIV measurements (from E.), interpolated onto the simulation grid by bilinear interpolation, with NaN cells (e.g. inside or very near the aggregate) filled by nearest-neighbor extrapolation. For the no-flow condition U = 0 and the equation reduces to Laplace’s equation (pure diffusion).

\subsection{Measuring chemotactic turning response from individual tracks}
The instantaneous swimming acceleration $\dot{\mathbf{v}}$ is estimated by applying a Savitzky–Golay filter to each velocity component with a window of 5 frames (1.0 s at 5 fps) and polynomial order 3. This provides a smoothed derivative that suppresses high-frequency noise while preserving the chemotaxis driven orientation dynamics. The two edge frames at each end of every track (half-window $= 2$) are excluded as $\dot{\mathbf{v}}$ cannot be computed there. Since $\gamma$ in Eqn. \eqref{eqngamma} diverges as $\delta \to 0$, steps where $\sin^2\theta < 0.05$, corresponding to the swimmer’s orientation lying within $\approx 13^{0}$ of the chemical gradient direction, are excluded.
Because noise is assumed uncorrelated over successive frames, the noise contribution to $\gamma$ averages to zero over the steps of a single track. We therefore use the per-track mean $\langle\gamma\rangle$ as the estimator of the deterministic chemotactic coupling, with within-track standard error $\sigma_k/\sqrt{n_k}$ quantifying the residual uncertainty from finite noise averaging ($n_k$ = number of valid steps in track $k$). The per-track standard deviation $\sigma_k$ estimates the amplitude of the stochastic reorientation noise, and the ratio $\sigma_k/|\langle\gamma_k \rangle|$ provides a per-track signal-to-noise ratio between stochastic and deterministic orientation dynamics. Before computing population-level statistics, track means $\langle\gamma\rangle_k$ are screened for outliers using Tukey’s fences ($1.5\times$ IQR criterion). Population means are reported as mean $\pm$ standard error of the mean across tracks in Fig. \ref{Figure3}(g).

\section*{Author Contributions} MP and RC designed the research. MP developed and optimized the staining and imaging protocol. RC and MP collected the microbial motility data. RC developed the data analysis pipeline and performed the analysis with key inputs from MP.  RC and MP constructed the theoretical framework for chemotaxis and transport dynamics. RC and MP wrote the manuscript. 
\section*{Acknowledgements} We thank the captain and crew of the RV Kilo Moana as well as the entire VIISTA science team, especially Kay D. Bidle and Kim Thamatrakoln for execution of the expedition at large. We thank H. F. Fredricks and the entire Benjamin van Mooy Lab at WHOI for expertise and execution of sediment trap collections for marine snow capture. We thank Qing Zhang for help with long-term imaging in Leg2 of the expedition. We thank Prakash Lab for vibrant discussions during group meetings, especially Ian Ho for fruitful debates on parameterizing microbial motility. We especially thank Hongquan Li for technical support with SQUID imaging. We thank the entire GCR research team for engaging discussions. RC acknowledges support from the Human Frontier Science Foundation. MP acknowledges funding support from Moore foundation, NSF GCR, Schmidt foundation and ARIA.

\bibliography{bacterial_hive}

@article{ALLDREDGE198841,
title = {Characteristics, dynamics and significance of marine snow},
journal = {Progress in Oceanography},
volume = {20},
number = {1},
pages = {41-82},
year = {1988},
issn = {0079-6611},
doi = {https://doi.org/10.1016/0079-6611(88)90053-5},
url = {https://www.sciencedirect.com/science/article/pii/0079661188900535},
author = {Alice L. Alldredge and Mary W. Silver}
}

@article{Azam2007,
  title = {Microbial structuring of marine ecosystems},
  volume = {5},
  ISSN = {1740-1534},
  url = {http://dx.doi.org/10.1038/nrmicro1747},
  DOI = {10.1038/nrmicro1747},
  number = {10},
  journal = {Nature Reviews Microbiology},
  publisher = {Springer Science and Business Media LLC},
  author = {Azam,  Farooq and Malfatti,  Francesca},
  year = {2007},
  month = oct,
  pages = {782–791}
}

@article{Stocker2012,
  title = {Ecology and Physics of Bacterial Chemotaxis in the Ocean},
  volume = {76},
  ISSN = {1098-5557},
  url = {http://dx.doi.org/10.1128/MMBR.00029-12},
  DOI = {10.1128/mmbr.00029-12},
  number = {4},
  journal = {Microbiology and Molecular Biology Reviews},
  publisher = {American Society for Microbiology},
  author = {Stocker,  Roman and Seymour,  Justin R.},
  year = {2012},
  month = dec,
  pages = {792–812}
}

@article{Pedley1992,
   author = "Pedley, T. J. and Kessler, J. O.",
   title = "Hydrodynamic Phenomena in Suspensions of Swimming Microorganisms", 
   journal= "Annual Review of Fluid Mechanics",
   year = "1992",
   volume = "24",
   number = "Volume 24, 1992",
   pages = "313-358",
   doi = "https://doi.org/10.1146/annurev.fl.24.010192.001525",
   url = "https://www.annualreviews.org/content/journals/10.1146/annurev.fl.24.010192.001525",
   publisher = "Annual Reviews",
   issn = "1545-4479",
   type = "Journal Article",
  }

@article{Raina2022,
  title = {Chemotaxis shapes the microscale organization of the ocean’s microbiome},
  volume = {605},
  ISSN = {1476-4687},
  url = {http://dx.doi.org/10.1038/s41586-022-04614-3},
  DOI = {10.1038/s41586-022-04614-3},
  number = {7908},
  journal = {Nature},
  publisher = {Springer Science and Business Media LLC},
  author = {Raina,  Jean-Baptiste et al.},
  year = {2022},
  month = apr,
  pages = {132–138}
}

@article{
Tara2015,
author = {Shinichi Sunagawa  and Luis Pedro Coelho  and Samuel Chaffron  and Jens Roat Kultima  and Karine Labadie  and Guillem Salazar  and Bardya Djahanschiri  and Georg Zeller  and Daniel R. Mende  and Adriana Alberti  and Francisco M. Cornejo-Castillo  and Paul I. Costea  and Corinne Cruaud  and Francesco d'Ovidio  and Stefan Engelen  and Isabel Ferrera  and Josep M. Gasol  and Lionel Guidi  and Falk Hildebrand  and Florian Kokoszka  and Cyrille Lepoivre  and Gipsi Lima-Mendez  and Julie Poulain  and Bonnie T. Poulos  and Marta Royo-Llonch  and Hugo Sarmento  and Sara Vieira-Silva  and Céline Dimier  and Marc Picheral  and Sarah Searson  and Stefanie Kandels-Lewis  and Tara Oceans coordinators and Chris Bowler  and Colomban de Vargas  and Gabriel Gorsky  and Nigel Grimsley  and Pascal Hingamp  and Daniele Iudicone  and Olivier Jaillon  and Fabrice Not  and Hiroyuki Ogata  and Stephane Pesant  and Sabrina Speich  and Lars Stemmann  and Matthew B. Sullivan  and Jean Weissenbach  and Patrick Wincker  and Eric Karsenti  and Jeroen Raes  and Silvia G. Acinas  and Peer Bork  and Emmanuel Boss  and Chris Bowler  and Michael Follows  and Lee Karp-Boss  and Uros Krzic  and Emmanuel G. Reynaud  and Christian Sardet  and Mike Sieracki  and Didier Velayoudon },
title = {Structure and function of the global ocean microbiome},
journal = {Science},
volume = {348},
number = {6237},
pages = {1261359},
year = {2015},
doi = {10.1126/science.1261359},
URL = {https://www.science.org/doi/abs/10.1126/science.1261359},
eprint = {https://www.science.org/doi/pdf/10.1126/science.1261359}}

@article{Ben2015,
author = {Collins, James R. and Edwards, Bethanie R. and Thamatrakoln, Kimberlee and Ossolinski, Justin E. and DiTullio, Giacomo R. and Bidle, Kay D. and Doney, Scott C. and Van Mooy, Benjamin A. S.},
title = {The multiple fates of sinking particles in the North Atlantic Ocean},
journal = {Global Biogeochemical Cycles},
volume = {29},
number = {9},
pages = {1471-1494},
year = {2015}
}

@article{Jeffery1922,
author = {Jeffery, George Barker  and Filon, Louis Napoleon George },
title = {The motion of ellipsoidal particles immersed in a viscous fluid},
journal = {Proceedings of the Royal Society of London. Series A, Containing Papers of a Mathematical and Physical Character},
volume = {102},
number = {715},
pages = {161-179},
year = {1922},
doi = {10.1098/rspa.1922.0078}}

@article{Stark2016,
  title = {Swimming in external fields},
  author = {Stark, Holger},
  journal = {The European Physical Journal Special Topics},
  volume = {225},
  issue = {11},
  pages = {2369–2387},
  year = {2016},
  publisher = {Springer},
  doi = {10.1140/epjst/e2016-60060-2},
  url = {https://doi.org/10.1140/epjst/e2016-60060-2}
}

@article{Suropriya2014,
  title = {Clusters, asters, and collective oscillations in chemotactic colloids},
  author = {Saha, Suropriya and Golestanian, Ramin and Ramaswamy, Sriram},
  journal = {Phys. Rev. E},
  volume = {89},
  issue = {6},
  pages = {062316},
  numpages = {7},
  year = {2014},
  month = {Jun},
  publisher = {American Physical Society},
  doi = {10.1103/PhysRevE.89.062316},
  url = {https://link.aps.org/doi/10.1103/PhysRevE.89.062316}
}

@book{leal2007advanced,
  title={Advanced Transport Phenomena: Fluid Mechanics and Convective Transport Processes},
  author={Leal, L. Gary},
  isbn={9780521849104},
  lccn={2007008779},
  series={Cambridge Series in Chemical Engineering},
  volume={7},
  year={2007},
  publisher={Cambridge University Press},
  address={Cambridge, UK}
}

@article{Stocker2016,
author = {Kwangmin Son  and Filippo Menolascina  and Roman Stocker },
title = {Speed-dependent chemotactic precision in marine bacteria},
journal = {Proceedings of the National Academy of Sciences},
volume = {113},
number = {31},
pages = {8624-8629},
year = {2016},
doi = {10.1073/pnas.1602307113},
URL = {https://www.pnas.org/doi/abs/10.1073/pnas.1602307113},
eprint = {https://www.pnas.org/doi/pdf/10.1073/pnas.1602307113}}

@article{Giering2017,
author = {Giering, Sarah L. C. and Sanders, Richard and Martin, Adrian P. and Henson, Stephanie A. and Riley, Jennifer S. and Marsay, Chris M. and Johns, David G.},
title = {Particle flux in the oceans: Challenging the steady state assumption},
journal = {Global Biogeochemical Cycles},
volume = {31},
number = {1},
pages = {159-171},
doi = {https://doi.org/10.1002/2016GB005424},
url = {https://agupubs.onlinelibrary.wiley.com/doi/abs/10.1002/2016GB005424},
eprint = {https://agupubs.onlinelibrary.wiley.com/doi/pdf/10.1002/2016GB005424},
year = {2017}
}

@article{barbara2003bacterial,
  title={Bacterial tracking of motile algae},
  author={Barbara, Greg M and Mitchell, James G},
  journal={FEMS Microbiology Ecology},
  volume={44},
  number={1},
  pages={79--87},
  year={2003},
  publisher={Oxford University Press},
  doi={10.1016/S0168-6496(03)00010-4}
}

@article{brumley2020cutting,
  title={Cutting Through the Noise: Bacterial Chemotaxis in Marine Microenvironments},
  author={Brumley, Douglas R and Carrara, Francesco and Hein, Andrew M and Hagstrom, George I and Levin, Simon A and Stocker, Roman},
  journal={Frontiers in Marine Science},
  volume={7},
  pages={527},
  year={2020},
  publisher={Frontiers Media SA},
  doi={10.3389/fmars.2020.00527}
}

@article{luo2022diversity,
  title={Diversity and origins of bacterial and archaeal viruses on sinking particles reaching the abyssal ocean},
  author={Luo, Elaine and Leu, Andy O and Eppley, John M and Karl, David M and Delong, Edward F},
  journal={The ISME Journal},
  volume={16},
  number={6},
  pages={1627--1635},
  year={2022},
  publisher={Oxford University Press},
  doi={10.1038/s41396-022-01202-1}
}

@article{durkin2022tracing,
  title={Tracing the path of carbon export in the ocean though DNA sequencing of individual sinking particles},
  author={Durkin, Colleen A and Cetini{\'c}, Ivona and Estapa, Margaret and Ljube{\v{s}}i{\'c}, Zrinka and Mucko, Maja and Neeley, Aimee and Omand, Melissa},
  journal={The ISME Journal},
  volume={16},
  number={8},
  pages={1896--1906},
  year={2022},
  publisher={Oxford University Press},
  doi={10.1038/s41396-022-01239-2}
}

@article{smith1992intense,
  title={Intense hydrolytic enzyme activity on marine aggregates and implications for rapid particle dissolution},
  author={Smith, David C and Simon, Meinhard Corporate and Alldredge, Alice L and Azam, Farooq},
  journal={Nature},
  volume={359},
  number={6391},
  pages={139--142},
  year={1992},
  publisher={Nature Publishing Group},
  doi={10.1038/359139a0}
}

@article{johansen2002variability,
  title={Variability in motility characteristics among marine bacteria},
  author={Johansen, Jens Efsen and Pinhassi, Jarone and Blackburn, Nicholas and Zweifel, Ulla Li and Hagstr{\"o}m, {\AA}ke},
  journal={Aquatic Microbial Ecology},
  volume={28},
  number={3},
  pages={229--237},
  year={2002},
  publisher={Inter-Research},
  doi={10.3354/ame028229}
}

@article{KiorboeJackson2001MarineSnow,
  author  = {Ki{\o}rboe, Thomas and Jackson, George A.},
  title   = {Marine Snow, Organic Solute Plumes, and Optimal
             Chemosensory Behavior of Bacteria},
  journal = {Limnology and Oceanography},
  year    = {2001},
  volume  = {46},
  number  = {6},
  pages   = {1309--1318},
  doi     = {10.4319/lo.2001.46.6.1309}
}

@article{Mitchell1995NaturalAssemblages,
  author  = {Mitchell, James G. and Pearson, Lynette and
             Dillon, Simon and Kantalis, Katerina},
  title   = {Natural Assemblages of Marine Bacteria Exhibiting
             High-Speed Motility and Large Accelerations},
  journal = {Applied and Environmental Microbiology},
  year    = {1995},
  volume  = {61},
  number  = {12},
  pages   = {4436--4440},
  doi     = {10.1128/aem.61.12.4436-4440.1995}
}

@article{Mitchell1991CellSize,
  author  = {Mitchell, James G.},
  title   = {The Influence of Cell Size on Marine Bacterial Motility and Energetics},
  journal = {Microbial Ecology},
  year    = {1991},
  volume  = {22},
  number  = {1},
  pages   = {227--238},
  doi     = {10.1007/BF02540225}
}

@article{Alcolombri2021,
  author    = {Uria Alcolombri and Fran{\c{c}}ois J. Peaudecerf and Vicente I. Fernandez and Lars Behrendt and Kang Soo Lee and Roman Stocker},
  title     = {Sinking enhances the degradation of organic particles by marine bacteria},
  journal   = {Nature Geoscience},
  volume    = {14},
  number    = {10},
  pages     = {775--780},
  year      = {2021},
  publisher = {Springer Science and Business Media LLC},
  doi       = {10.1038/s41561-021-00817-x}
}

@article{Boyd2019,
  doi = {10.1038/s41586-019-1098-2},
  url = {https://doi.org/10.1038/s41586-019-1098-2},
  year = {2019},
  month = apr,
  publisher = {Springer Science and Business Media {LLC}},
  volume = {568},
  number = {7752},
  pages = {327--335},
  author = {Philip W. Boyd and Herv{\'{e}} Claustre and Marina Levy and David A. Siegel and Thomas Weber},
  title = {Multi-faceted particle pumps drive carbon sequestration in the ocean},
  journal = {Nature}
}

@article{Ducklow2001,
  doi = {10.5670/oceanog.2001.06},
  url = {https://doi.org/10.5670/oceanog.2001.06},
  year = {2001},
  publisher = {The Oceanography Society},
  volume = {14},
  number = {4},
  pages = {50--58},
  author = {Hugh Ducklow and Deborah Steinberg and Ken Buesseler},
  title = {Upper Ocean Carbon Export and the Biological Pump},
  journal = {Oceanography}
}

@article{chajwa2024hidden,
  author    = {Chajwa, Rahul and Flaum, Eliott and Bidle, Kay D. and Van Mooy, Benjamin and Prakash, Manu},
  title     = {Hidden comet tails of marine snow impede ocean-based carbon sequestration},
  journal   = {Science},
  volume    = {386},
  number    = {6718},
  pages     = {eadl5767},
  year      = {2024},
  doi       = {10.1126/science.adl5767},
  url       = {https://www.science.org/doi/10.1126/science.adl5767}
}

@book{kiorboe2008mechanistic,
  title     = {A Mechanistic Approach to Plankton Ecology},
  author    = {Ki{\o}rboe, Thomas},
  year      = {2008},
  publisher = {Princeton University Press},
  address   = {Princeton, NJ},
  isbn      = {978-0-691-13422-2},
  pages     = {209}
}

@article{Azam1983Ecological,
  author    = {Azam, F. and Fenchel, T. and Field, J. G. and Gray, J. S. and Meyer-Reil, L. A. and Thingstad, F.},
  title     = {The Ecological Role of Water-Column Microbes in the Sea},
  journal   = {Marine Ecology Progress Series},
  volume    = {10},
  pages     = {257--263},
  year      = {1983},
  doi       = {10.3354/meps010257}
}

@article{Berg1977,
  author    = {Berg, H. C. and Purcell, E. M.},
  title     = {Physics of Chemoreception},
  journal   = {Biophysical Journal},
  volume    = {20},
  number    = {2},
  pages     = {193--219},
  year      = {1977},
  month     = {November},
  doi       = {10.1016/S0006-3495(77)85544-6},
  pmid      = {911982},
  pmcid     = {PMC1473391}
}

@book{berg2004ecoli,
  author    = {Howard C. Berg},
  title     = {E. coli in Motion},
  publisher = {Springer Science \& Business Media},
  year      = {2004},
  address   = {New York},
  isbn      = {978-0-387-00888-2},
  doi       = {10.1007/b97370}
}

@article{Kiorboe2002Colonization,
  author  = {Ki{\o}rboe, Thomas and Grossart, Hans-Peter and
             Ploug, Helle and Tang, Kam W.},
  title   = {Mechanisms and Rates of Bacterial Colonization of
             Sinking Aggregates},
  journal = {Applied and Environmental Microbiology},
  year    = {2002},
  volume  = {68},
  number  = {8},
  pages   = {3996--4006},
  doi     = {10.1128/AEM.68.8.3996-4006.2002}
}

@article{Desai2019HydrodynamicColonization,
  author  = {Desai, Nikhil and Shaik, Vaseem A. and Ardekani, Arezoo M.},
  title   = {Hydrodynamic Interaction Enhances Colonization of Sinking
             Nutrient Sources by Motile Microorganisms},
  journal = {Frontiers in Microbiology},
  year    = {2019},
  volume  = {10},
  pages   = {289},
  doi     = {10.3389/fmicb.2019.00289}
}

@article{StaleyKonopka1985,
  author  = {Staley, James T. and Konopka, Allan},
  title   = {Measurement of In Situ Activities of Nonphotosynthetic
             Microorganisms in Aquatic and Terrestrial Habitats},
  journal = {Annual Review of Microbiology},
  year    = {1985},
  volume  = {39},
  pages   = {321--346},
  doi     = {10.1146/annurev.mi.39.100185.001541}
}

@article{Lambert2017ISCA,
  author  = {Lambert, Bennett S. and Raina, Jean-Baptiste and
             Fernandez, Vicente I. and Rinke, Christian and
             Siboni, Nachshon and Rubino, Francesco and
             Hugenholtz, Philip and Tyson, Gene W. and
             Seymour, Justin R. and Stocker, Roman},
  title   = {A microfluidics-based in situ chemotaxis assay to study
             the behaviour of aquatic microbial communities},
  journal = {Nature Microbiology},
  year    = {2017},
  volume  = {2},
  number  = {10},
  pages   = {1344--1349},
  doi     = {10.1038/s41564-017-0010-9}
}

@article{grossart2001bacterial,
  title={Bacterial motility in the sea and its ecological implications},
  author={Grossart, Hans-Peter and Riemann, Lasse and Azam, Farooq},
  journal={Aquatic Microbial Ecology},
  volume={25},
  number={3},
  pages={247--258},
  year={2001},
  doi={10.3354/ame025247},
  url={https://doi.org}
}

@article{Locsei2009,
  author     = {Locsei, J. T. and Pedley, T. J.},
  title      = {Run and Tumble Chemotaxis in a Shear Flow: The Effect of Temporal Comparisons, Persistence, Rotational Diffusion, and Cell Shape},
  journal    = {Bulletin of Mathematical Biology},
  volume     = {71},
  number     = {5},
  pages      = {1089--1116},
  year       = {2009},
  month     = {Jul},
  doi        = {10.1007/s11538-009-9395-9},
  url        = {https://doi.org/10.1007/s11538-009-9395-9},
  publisher = {Springer Science and Business Media LLC}
}

@article{Stephens2024DirectObservations,
  author  = {Stephens, Brandon M. and Durkin, Colleen A. and Sharpe, Garrett and Nguyen, Trang T. H. and Albers, Justine and Estapa, Margaret L. and Steinberg, Deborah K. and Levine, Naomi M. and Gifford, Scott M. and Carlson, Craig A. and Boyd, Philip W. and Santoro, Alyson E.},
  title   = {Direct observations of microbial community succession on sinking marine particles},
  journal = {The ISME Journal},
  year    = {2024},
  volume  = {18},
  number  = {1},
  pages   = {wrad010},
  doi     = {10.1093/ismejo/wrad010}
}

@article{Pelve2017BacterialSuccession,
  author  = {Pelve, Erik A. and Fontanez, Kristina M. and DeLong, Edward F.},
  title   = {Bacterial Succession on Sinking Particles in the Ocean's Interior},
  journal = {Frontiers in Microbiology},
  year    = {2017},
  volume  = {8},
  pages   = {2269},
  doi     = {10.3389/fmicb.2017.02269}
}

@article{Turczynowicz_2026,
  title={Bridging advection and diffusion in the encounter dynamics of sedimenting marine snow},
  volume={1031},
  DOI={10.1017/jfm.2026.11282},
  art_number={A5},
  journal={Journal of Fluid Mechanics},
  publisher={Cambridge University Press},
  author={Turczynowicz, J. and Waszkiewicz, R. and Słomka, J. and Lisicki, M.},
  year={2026}
}

@article{Durkin_2026,
  title={The Biology of Marine Snowflakes},
  volume={18},
  DOI={10.1146/annurev-marine-040523-021832},
  number={1},
  journal={Annual Review of Marine Science},
  publisher={Annual Reviews},
  author={Durkin, Colleen A.},
  year={2026},
  pages={327--350}
}

@article{Slomka2023EncounterRates,
  author  = {S{\l}omka, Jonasz and Alcolombri, Uria and
             Carrara, Francesco and Foffi, Riccardo and
             Peaudecerf, Fran{\c c}ois J. and Zbinden, Matti
             and Stocker, Roman},
  title   = {Encounter rates prime interactions between microorganisms},
  journal = {Interface Focus},
  year    = {2023},
  volume  = {13},
  number  = {2},
  pages   = {20220059},
  doi     = {10.1098/rsfs.2022.0059}
}

@article{Keegstra2025,
  author = {Keegstra, Johannes M. and Landry, Zachary C. and Zweifel, Sophie T. and Roller, Benjamin R. K. and Baumgartner, Dieter A. and Carrara, Francesco and Mart{\'i}nez-P{\'e}rez, Clara and Clerc, Estelle E. and Ackermann, Martin and Stocker, Roman},
  title = {Risk--reward trade-off during carbon starvation generates dichotomy in motility endurance among marine bacteria},
  journal = {Nature Microbiology},
  volume = {10},
  number = {6},
  pages = {1393--1403},
  year = {2025},
  doi = {10.1038/s41564-025-01997-7},
  url = {https://nature.com}
}

@article{blackburn1998microscale,
  author    = {Blackburn, Nick and Fenchel, Tom and Mitchell, James G.},
  title     = {Microscale Nutrient Patches in Planktonic Habitats Shown by Chemotactic Bacteria},
  journal   = {Science},
  volume    = {282},
  number    = {5397},
  pages     = {2254--2256},
  year      = {1998},
  doi       = {10.1126/science.282.5397.2254},
  url       = {https://www.science.org/doi/10.1126/science.282.5397.2254}
}

@article{Slomka2020,
  author  = {S{\l}omka, Jonasz and Alcolombri, Uria and Secchi, Eleonora and Stocker, Roman and Fernandez, Vicente I.},
  title   = {Encounter rates between bacteria and small sinking particles},
  journal = {New Journal of Physics},
  volume  = {22},
  number  = {4},
  pages   = {043016},
  year    = {2020},
  doi     = {10.1088/1367-2630/ab73c9}
}

@article{Joshi2025,
  author  = {Joshi, Harshit and Chajwa, Rahul and Ramaswamy, Sriram and Menon, Narayanan and Govindarajan, Rama},
  title   = {Dynamics and clustering of sedimenting disc lattices},
  journal = {Journal of Fluid Mechanics},
  volume  = {1017},
  pages   = {A1},
  year    = {2025},
  doi     = {10.1017/jfm.2025.10467}
}

@article{Borer2022PorousMS,
  author    = {Benedict Borer and Irene H. Zhang and Amy E. Baker and George A. O'Toole and Andrew R. Babbin},
  title     = {Porous marine snow differentially benefits chemotactic, motile, and nonmotile bacteria},
  journal   = {PNAS Nexus},
  volume    = {2},
  number    = {2},
  year      = {2023},
  pages     = {pgac311},
  doi       = {10.1093/pnasnexus/pgac311},
  url       = {https://academic.oup.com/pnasnexus/article/2/2/pgac311/6961070}
}

@article{
Stocker2008,
author = {Roman Stocker  and Justin R. Seymour  and Azadeh Samadani  and Dana E. Hunt  and Martin F. Polz },
title = {Rapid chemotactic response enables marine bacteria to exploit ephemeral microscale nutrient patches},
journal = {Proceedings of the National Academy of Sciences},
volume = {105},
number = {11},
pages = {4209-4214},
year = {2008},
doi = {10.1073/pnas.0709765105},
URL = {https://www.pnas.org/doi/abs/10.1073/pnas.0709765105},
eprint = {https://www.pnas.org/doi/pdf/10.1073/pnas.0709765105}}

@article{Durham2013,
  author    = {Durham, William M. and Climent, Eric and Barry, Michael and De Lillo, Filippo and Boffetta, Guido and Cencini, Massimo and Stocker, Roman},
  title     = {Turbulence drives microscale patches of motile phytoplankton},
  journal   = {Nature Communications},
  volume    = {4},
  pages     = {2148},
  year      = {2013},
  doi       = {10.1038/ncomms3148},
  issn      = {2041-1723},
  url       = {https://nature.com}
}

@article {Li2020Squid,
	author = {Li, Hongquan and Krishnamurthy, Deepak and Li, Ethan and Vyas, Pranav and Akireddy, Niranjan and Chai, Corena and Prakash, Manu},
	title = {Squid: Simplifying Quantitative Imaging Platform Development and Deployment},
	elocation-id = {2020.12.28.424613},
	year = {2020},
	doi = {10.1101/2020.12.28.424613},
	publisher = {Cold Spring Harbor Laboratory},
	URL = {https://biorxiv.org},
	eprint = {https://biorxiv.org.full.pdf},
	journal = {bioRxiv}
}

@incollection{Ducklow2001b,
  author    = {Ducklow, Hugh W.},
  title     = {Bacterioplankton},
  booktitle = {Encyclopedia of Ocean Sciences},
  editor    = {Steele, John H. and Turekian, Karl K. and Thorpe, Steve A.},
  publisher = {Academic Press},
  year      = {2001},
  pages     = {217--224}
}

@article{Locsei2009RunAT,
  title     = {Run and Tumble Chemotaxis in a Shear Flow: The Effect of Temporal Comparisons, Persistence, Rotational Diffusion, and Cell Shape},
  author    = {Janos Tobias Locsei and Timothy J. Pedley},
  journal   = {Bulletin of Mathematical Biology},
  year      = {2009},
  volume    = {71},
  number    = {5},
  pages     = {1089--1116},
  doi       = {10.1007/s11538-009-9395-9},
  url       = {https://link.springer.com/article/10.1007/s11538-009-9395-9}
}

@article{Ferguson1984,
  author  = {Ferguson, R. L. and Buckley, E. N. and Palumbo, A. V.},
  title   = {Response of marine bacterioplankton to differential filtration and confinement},
  journal = {Applied and Environmental Microbiology},
  volume  = {47},
  number  = {1},
  pages   = {49--55},
  year    = {1984}
}

@article{MunsonMcGee2022,
  author    = {Munson-McGee, Jacob H. and Lindsay, M. Lindsay and others and Stepanauskas, Ramunas},
  title     = {Decoupling of respiration rates and abundance in marine prokaryoplankton},
  journal   = {Nature},
  volume    = {612},
  number    = {7941},
  pages     = {764--770},
  year      = {2022},
  doi       = {10.1038/s41586-022-05505-3},
  url       = {https://www.nature.com/articles/s41586-022-05505-3}
}

@article{ploug1999photosynthesis,
  title={Photosynthesis, respiration, and carbon turnover in sinking marine snow from surface waters of Southern California Bight: implications for the carbon cycle in the ocean},
  author={Ploug, H. and Grossart, H. P. and Azam, F. and J{\o}rgensen, B. B.},
  journal={Marine Ecology Progress Series},
  volume={179},
  pages={1--11},
  year={1999},
  publisher={Inter-Research},
  url={https://www.int-res.com/journals/meps/articles/meps179001}
}

@article{Kiorboe2003,
  author  = {Ki{\o}rboe, Thomas and Tang, Kam and Grossart, Hans-Peter and Ploug, Helle},
  title   = {Dynamics of Microbial Communities on Marine Snow Aggregates: Colonization, Growth, Detachment, and Grazing Mortality of Attached Bacteria},
  journal = {Applied and Environmental Microbiology},
  year    = {2003},
  volume  = {69},
  number  = {6},
  pages   = {3036--3047},
  doi     = {10.1128/AEM.69.6.3036-3047.2003}
}

@article{Kiorboe2001FluidMotion,
  author    = {Ki{\o}rboe, Thomas and Ploug, Helle and Thygesen, Uffe H{\o}gsbro},
  title     = {Fluid motion and solute distribution around sinking aggregates. {I}. {Small-scale} fluxes and heterogeneity of nutrients in the pelagic environment},
  journal   = {Marine Ecology Progress Series},
  volume    = {211},
  pages     = {1--13},
  year      = {2001},
  doi       = {10.3354/meps211001},
  url       = {https://www.int-res.com/journals/meps/articles/meps211001}
}

@article{Peterson2005,
author = {Peterson, Michael L. and Wakeham, Stuart G. and Lee, Cindy and Askea, Meaghan A. and Miquel, Juan Carlos},
title = {Novel techniques for collection of sinking particles in the ocean and determining their settling rates},
journal = {Limnology and Oceanography: Methods},
volume = {3},
number = {12},
pages = {520-532},
doi = {https://doi.org/10.4319/lom.2005.3.520},
url = {https://aslopubs.onlinelibrary.wiley.com/doi/abs/10.4319/lom.2005.3.520},
eprint = {https://aslopubs.onlinelibrary.wiley.com/doi/pdf/10.4319/lom.2005.3.520},
year = {2005}
}

@article{Lamborg2008Flux,
  author    = {Lamborg, C. H. and Buesseler, K. O. and Valdes, J. and Bertrand, C. H. and Bidigare, R. and Manganini, S. and Pike, S. and Steinberg, D. and Trull, T. and Wilson, S. E.},
  title     = {The flux of bio- and lithogenic material associated with sinking particles in the mesopelagic ``twilight zone'' of the northwest and North Central Pacific Ocean},
  journal   = {Deep Sea Research Part II: Topical Studies in Oceanography},
  volume    = {55},
  number    = {14--15},
  pages     = {1540--1563},
  year      = {2008},
  doi       = {10.1016/j.dsr2.2008.04.011}
}

@article{IversenPloug2010,
  author  = {Iversen, M. H. and Ploug, H.},
  title   = {Ballast minerals and the sinking carbon flux in the ocean:
             carbon-specific respiration rates and sinking velocity
             of marine snow aggregates},
  journal = {Biogeosciences},
  year    = {2010},
  volume  = {7},
  number  = {9},
  pages   = {2613--2624},
  doi     = {10.5194/bg-7-2613-2010}
}

@misc{chajwa_activeparticles_2026,
  author       = {Chajwa, Rahul},
  title        = {rahulchajwa/activeparticles-in-turbulence: Active-Particles-In-Turbulence},
  month        = mar,
  year         = {2026},
  publisher    = {Zenodo},
  version      = {v1.0.0},
  doi          = {10.5281/zenodo.18882257},
  url          = {https://doi.org/10.5281/zenodo.18882257}
}

@article{Son2013,
  author    = {Son, Kwangmin and Guasto, Jeffrey S. and Stocker, Roman},
  title     = {Bacteria can exploit a flagellar buckling instability to change direction},
  journal   = {Nature Physics},
  volume    = {9},
  number    = {8},
  pages     = {494--498},
  year      = {2013},
  doi       = {10.1038/nphys2676},
  url       = {https://nature.com}
}

@article{purcell1977life,
  title={Life at low Reynolds number},
  author={Purcell, Edward M},
  journal={American Journal of Physics},
  volume={45},
  number={1},
  pages={3--11},
  year={1977},
  publisher={American Association of Physics Teachers}
}

@article{makarieva2008mean,
  title={Mean mass-specific metabolic rates are strikingly similar across life's major kingdoms},
  author={Makarieva, Anastassia M and Gorshkov, Victor G and Li, Bai-Lian and Chown, Steven L and Reich, Peter B and Gavrilov, Valery M},
  journal={Proceedings of the National Academy of Sciences},
  volume={105},
  number={44},
  pages={16994--16999},
  year={2008},
  publisher={National Academies of Sciences}
}

@article{Maitra2014,
  title = {Activating Membranes},
  author = {Maitra, Ananyo and Srivastava, Pragya and Rao, Madan and Ramaswamy, Sriram},
  journal = {Phys. Rev. Lett.},
  volume = {112},
  issue = {25},
  pages = {258101},
  numpages = {5},
  year = {2014},
  month = {Jun},
  publisher = {American Physical Society},
  doi = {10.1103/PhysRevLett.112.258101},
  url = {https://link.aps.org/doi/10.1103/PhysRevLett.112.258101}
}

@article{Brotto2013,
  title = {Hydrodynamics of Confined Active Fluids},
  author = {Brotto, Tommaso and Caussin, Jean-Baptiste and Lauga, Eric and Bartolo, Denis},
  journal = {Phys. Rev. Lett.},
  volume = {110},
  issue = {3},
  pages = {038101},
  numpages = {5},
  year = {2013},
  month = {Jan},
  publisher = {American Physical Society},
  doi = {10.1103/PhysRevLett.110.038101},
  url = {https://link.aps.org/doi/10.1103/PhysRevLett.110.038101}
}

@article{Cael2021,
author = {Cael, B. B. and Cavan, Emma L. and Britten, Gregory L.},
title = {Reconciling the Size-Dependence of Marine Particle Sinking Speed},
journal = {Geophysical Research Letters},
volume = {48},
number = {5},
pages = {e2020GL091771},
doi = {https://doi.org/10.1029/2020GL091771},
url = {https://agupubs.onlinelibrary.wiley.com/doi/abs/10.1029/2020GL091771},
eprint = {https://agupubs.onlinelibrary.wiley.com/doi/pdf/10.1029/2020GL091771},
note = {e2020GL091771 2020GL091771},
year = {2021}
}

@article{Seymour2010Chemoattraction,
  author    = {Seymour, Justin R. and Sim\'{o}, Rafel and Ahmed, Tanvir and Stocker, Roman},
  title     = {Chemoattraction to Dimethylsulfoniopropionate Throughout the Marine Microbial Food Web},
  journal   = {Science},
  volume    = {329},
  number    = {5989},
  pages     = {342--345},
  year      = {2010},
  doi       = {10.1126/science.1188418},
  URL       = {https://science.org}
}

@article{romanczuk2012active,
  title={Active Brownian particles: From individual to collective stochastic dynamics},
  author={Romanczuk, Pawel and B{\"a}r, Markus and Ebeling, Werner and Lindner, Benjamin and Schimansky-Geier, Lutz},
  journal={The European Physical Journal Special Topics},
  volume={202},
  number={1},
  pages={1--162},
  year={2012},
  publisher={Springer},
  doi={10.1140/epjst/e2012-01529-y}
}

@article{KellerSegel1971,
  author  = {Keller, Evelyn F. and Segel, Lee A.},
  title   = {Model for chemotaxis},
  journal = {Journal of Theoretical Biology},
  year    = {1971},
  volume  = {30},
  number  = {2},
  pages   = {225--234},
  doi     = {10.1016/0022-5193(71)90050-6}
}

@article{
Federico2009,
author = {Federico M. Lauro  and Diane McDougald  and Torsten Thomas  and Timothy J. Williams  and Suhelen Egan  and Scott Rice  and Matthew Z. DeMaere  and Lily Ting  and Haluk Ertan  and Justin Johnson  and Steven Ferriera  and Alla Lapidus  and Iain Anderson  and Nikos Kyrpides  and A. Christine Munk  and Chris Detter  and Cliff S. Han  and Mark V. Brown  and Frank T. Robb  and Staffan Kjelleberg  and Ricardo Cavicchioli },
title = {The genomic basis of trophic strategy in marine bacteria},
journal = {Proceedings of the National Academy of Sciences},
volume = {106},
number = {37},
pages = {15527-15533},
year = {2009},
doi = {10.1073/pnas.0903507106},
URL = {https://www.pnas.org/doi/abs/10.1073/pnas.0903507106},
eprint = {https://www.pnas.org/doi/pdf/10.1073/pnas.0903507106}}

@article{Giovannoni2005,
  author    = {Giovannoni, Stephen J. and Tripp, H. James and Givan, Scott and Podar, Mircea and Vergin, Kevin L. and Baptista, Damon and Bibbs, Lisa and Eads, Jonathan and Richardson, Toby H. and Noordewier, Michiel and Rapp{\'e}, Michael S. and Short, Jay M. and Carrington, James C. and Mathur, Eric J.},
  title     = {Genome streamlining in a cosmopolitan oceanic bacterium},
  journal   = {Science},
  volume    = {309},
  number    = {5738},
  pages     = {1242--1245},
  year      = {2005},
  doi       = {10.1126/science.1114057},
  url       = {https://science.org},
  issn      = {0036-8075},
  eprint    = {https://science.org}
}

@article{Hallstrom2022,
  author = {Hallstr{\o}m, S{\o}ren and Raina, Jean-Baptiste and Ostrowski, Martin and Parks, Donovan H. and Tyson, Gene W. and Hugenholtz, Philip and Stocker, Roman and Seymour, Justin R. and Riemann, Lasse},
  title = {Chemotaxis may assist marine heterotrophic bacterial diazotrophs to find microzones suitable for N$_2$ fixation in the pelagic ocean},
  journal = {The ISME Journal},
  volume = {16},
  number = {11},
  pages = {2525--2534},
  year = {2022},
  doi = {10.1038/s41539-022-00947-x}
}

@article{Plough2001,
author = {Ploug, Helle},
title = {Small-scale oxygen fluxes and remineralization in sinking aggregates},
journal = {Limnology and Oceanography},
volume = {46},
number = {7},
pages = {1624-1631},
doi = {https://doi.org/10.4319/lo.2001.46.7.1624},
url = {https://aslopubs.onlinelibrary.wiley.com/doi/abs/10.4319/lo.2001.46.7.1624},
eprint = {https://aslopubs.onlinelibrary.wiley.com/doi/pdf/10.4319/lo.2001.46.7.1624},
year = {2001}
}

@article{Bell1972,
author = {WAYNE BELL  and RALPH MITCHELL },
title = {CHEMOTACTIC AND GROWTH RESPONSES OF MARINE BACTERIA TO ALGAL EXTRACELLULAR PRODUCTS},
journal = {The Biological Bulletin},
volume = {143},
number = {2},
pages = {265-277},
year = {1972},
doi = {10.2307/1540052},

URL = {https://www.journals.uchicago.edu/doi/abs/10.2307/1540052},
eprint = {https://www.journals.uchicago.edu/doi/pdf/10.2307/1540052}
}

@article{Cael2024TinyCU,
  title   = {Tiny comets under the sea},
  author  = {B. B. Cael and Lionel Guidi},
  journal = {Science},
  year    = {2024},
  volume  = {386},
  number  = {6718},
  pages   = {149--150},
  doi     = {10.1126/science.ads5642}
}

@article{martin1987vertex,
  title={{VERTEX}: Carbon cycling in the northeast {Pacific}},
  author={Martin, John H. and Knauer, George A. and Karl, David M. and Broenkow, William W.},
  journal={Deep Sea Research Part A. Oceanographic Research Papers},
  volume={34},
  number={2},
  pages={267--285},
  year={1987},
  publisher={Elsevier},
  doi={10.1016/0198-0149(87)90086-0},
  url={https://doi.org/10.1016/0198-0149(87)90086-0}
}

@article{Thar_2003,
  author    = {Thar, Roland and Kühl, Michael},
  title     = {Bacteria are not too small for spatial sensing of chemical gradients: An experimental evidence},
  journal   = {Proceedings of the National Academy of Sciences},
  volume    = {100},
  number    = {10},
  pages     = {5748--5753},
  year      = {2003},
  month     = {apr},
  doi       = {10.1073/pnas.1030795100},
  url       = {https://www.pnas.org/doi/10.1073/pnas.1030795100}
}

@article{Thar2001,
  author    = {Thar, Roland and Fenchel, Tom},
  title     = {True Chemotaxis in Oxygen Gradients of the Sulfur-Oxidizing Bacterium \emph{Thiovulum majus}},
  journal   = {Applied and Environmental Microbiology},
  volume    = {67},
  number    = {7},
  pages     = {3299--3303},
  year      = {2001},
  publisher = {American Society for Microbiology},
  doi       = {10.1128/AEM.67.7.3299-3303.2001},
  url       = {https://journals.asm.org/doi/10.1128/AEM.67.7.3299-3303.2001}
}

@article{
Shelley2012,
author = {Leif Ristroph  and Matthew N. J. Moore  and Stephen Childress  and Michael J. Shelley  and Jun Zhang },
title = {Sculpting of an erodible body by flowing water},
journal = {Proceedings of the National Academy of Sciences},
volume = {109},
number = {48},
pages = {19606-19609},
year = {2012},
doi = {10.1073/pnas.1212286109},
URL = {https://www.pnas.org/doi/abs/10.1073/pnas.1212286109},
eprint = {https://www.pnas.org/doi/pdf/10.1073/pnas.1212286109}}

@article{Omand2020SinkingFO,
  author    = {Melissa M. Omand and Rama Govindarajan and Jing He and Amala Mahadevan},
  title     = {Sinking flux of particulate organic matter in the oceans: Sensitivity to particle characteristics},
  journal   = {Scientific Reports},
  volume    = {10},
  number    = {1},
  pages     = {5582},
  year      = {2020},
  doi       = {10.1038/s41598-020-60424-5},
  url       = {https://www.nature.com/articles/s41598-020-60424-5}
}

@article{ChajwaCaus2026,
  title = {Active caustics},
  author = {Chajwa, Rahul and Rajarshi, C. and Govindarajan, Rama and Ramaswamy, Sriram},
  journal = {Phys. Rev. Fluids},
  volume = {11},
  issue = {3},
  pages = {033104},
  numpages = {23},
  year = {2026},
  month = {Mar},
  publisher = {American Physical Society},
  doi = {10.1103/m3v2-bdvj},
  url = {https://link.aps.org/doi/10.1103/m3v2-bdvj}
}

@book{huang1987statistical,
  author    = {Huang, Kerson},
  title     = {Statistical Mechanics},
  edition   = {2nd},
  publisher = {John Wiley \& Sons},
  year      = {1987},
  address   = {New York, NY},
  isbn      = {9780471815181}
}

@article{peixoto2026global,
  title={Global change reshapes marine microbial communities, functions and One Health outcomes},
  author={Peixoto, Raquel S. and Voolstra, Christian R. and Duarte, G. and others},
  journal={Nature Reviews Microbiology},
  year={2026},
  publisher={Nature Publishing Group}
}

@article{Gray2019,
  author    = {Gray, William T. and Govers, Sander K. and Xiang, Yingjie and Parry, Bradley R. and Campos, Manuel and Kim, Sangjin and Jacobs-Wagner, Christine},
  title     = {Nucleoid Size Scaling and Intracellular Organization of Translation Across Bacteria},
  journal   = {Cell},
  volume    = {177},
  number    = {7},
  pages     = {1632--1648.e20},
  year      = {2019},
  doi       = {10.1016/j.cell.2019.05.017},
  pmid      = {31150626},
  pmcid     = {PMC6629263}
}

@article{taute2015high,
  title={High-throughput 3D tracking of bacteria on a standard phase contrast microscope},
  author={Taute, Katja M and Gude, Solenn and Tans, Sander J and Shimizu, Thomas S},
  journal={Nature Communications},
  volume={6},
  number={1},
  pages={8776},
  year={2015},
  publisher={Nature Publishing Group},
  doi={10.1038/ncomms9776},
  url={https://nature.com}
}

@article{Kim2024,
  author    = {Kim, Jihwan and Lee, Sang Joon},
  title     = {Digital in-line holographic microscopy for label-free identification and tracking of biological cells},
  journal   = {Military Medical Research},
  volume    = {11},
  number    = {1},
  pages     = {38},
  year      = {2024},
  doi       = {10.1186/s40779-024-00541-8},
  url       = {https://doi.org/10.1186/s40779-024-00541-8}
}

@article{Bloxham2026,
  author    = {Bloxham, Blox and Lee, Hyunseok and Gore, Jeff},
  title     = {Repulsion from slow-diffusing nutrients improves microbial chemotaxis towards moving sources},
  journal   = {Nature Communications},
  volume    = {17},
  number    = {1},
  pages     = {71148},
  year      = {2026},
  doi       = {10.1038/s41467-026-71148-x},
  url       = {https://www.nature.com/articles/s41467-026-71148-x}
}

@article{Kamp2024,
  author    = {Kamp, Marlous and Surmacki, Jakub and Segarra Mondejar, Marc and Young, Tom and Chrabaszcz, Kinga and Joud, Fadwa and Zecchini, Valerio and Speed, Amy and Frezza, Christian and Bohndiek, Sarah E.},
  title     = {Raman micro-spectroscopy reveals the spatial distribution of fumarate in cells and tissues},
  journal   = {Nature Communications},
  volume    = {15},
  pages     = {5386},
  year      = {2024},
  doi       = {10.1038/s41467-024-49403-w},
  url       = {https://nature.com}
}

@article{Crocker2025,
  author    = {Crocker, Kyle and Skwara, Abigail and Kannan, Rathi and Murugan, Arvind and Kuehn, Seppe},
  title     = {Microbial functional guilds respond cohesively to rapidly fluctuating environments},
  journal   = {bioRxiv},
  year      = {2025},
  doi       = {10.1101/2025.01.30.635766},
  url       = {https://doi.org}
}

@article{Jaqaman2008RobustSP,
  title     = {Robust single particle tracking in live cell time-lapse sequences},
  author    = {Khuloud Jaqaman and Dinah Loerke and Marcel Mettlen and Hirotaka Kuwata and Sergio Grinstein and Sandra L. Schmid and Gaudenz Danuser},
  journal   = {Nature Methods},
  volume    = {5},
  number    = {8},
  pages     = {695--702},
  year      = {2008},
  doi       = {10.1038/nmeth.1237},
  url       = {https://doi.org/10.1038/nmeth.1237}
}
\bibliographystyle{ieeetr}

\pagebreak
\clearpage
\newpage
\onecolumngrid
\section*{Supplementary Material for Chemotactic Horizons Regulate Threshold Dynamics in the Biological Carbon Pump}
\setcounter{figure}{0} 
\renewcommand{\thefigure}{S\arabic{figure}}
\section*{I. Error Analysis}

\subsection*{A. Background noise from ship}
All imaging presented in our work was done at sea on a table top experiment in a research vessel. Given that we are operating in a dilute limit of microbe number density, any anisotropic collective oscillations we see in flow datasets is an imposed noise from the surroundings. We observed residual noise in the non-motile tracks in our high resolution imaging datasets, termed `diffusive' in the paper. This random motion has three sources: 1) thermal noise, 2) standard active noise, and 3) ship vibrations. These distinct sources convolve to produce the high frequency noise signal manifest in the non-motile tracks. Below we characterize this noise. The anisotropy of vibrations strongly suggests ship vibrations to be a dominant source that perturbs the non-motile tracks but it's strength is too weak to affect the motile trajectories.

\begin{figure*}[h]
 \begin{center}
 \includegraphics[width=17.5 cm]{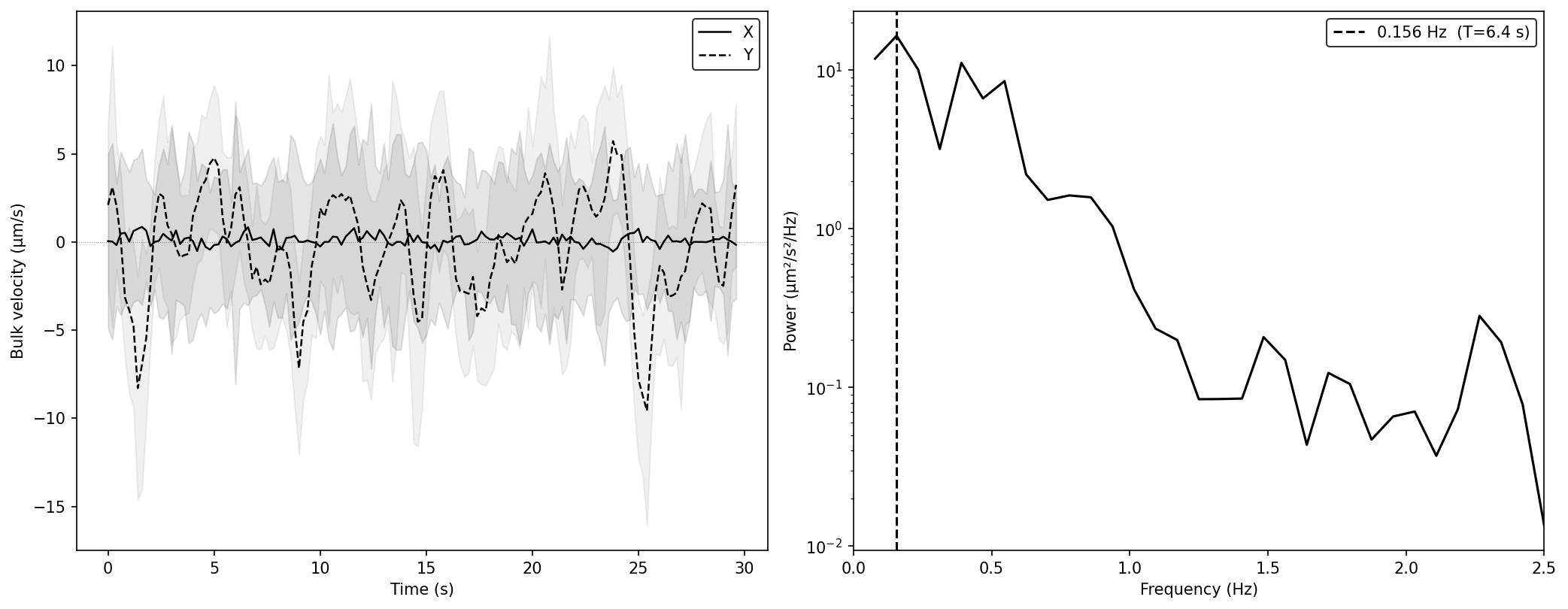}
 \caption{\textbf{Ship noise}: (i) (left) the mean velocity of all non-motile swimmer identified using KDE anlaysis, with the variance in the grey shaded region, shows directional vibrations along y (the channel's long axis), with horizontal vibrations relatively negligible. Note that all degrees of freedom of the ship manifest as anisotropic response of the channel because of channel geometry (ii)(right) The noise power spectrum of this background fluctuation peak at 0.15 Hz ( $2\pi/6.4 \mathrm{s}$), which corresponds roughly to the low frequency ship wobble.}
 \label{FigureS1}
 \end{center}
 \end{figure*}

\subsection*{B. Flow Measurement}
\begin{figure*}[ht]
 \begin{center}
 \includegraphics[width=17.5 cm]{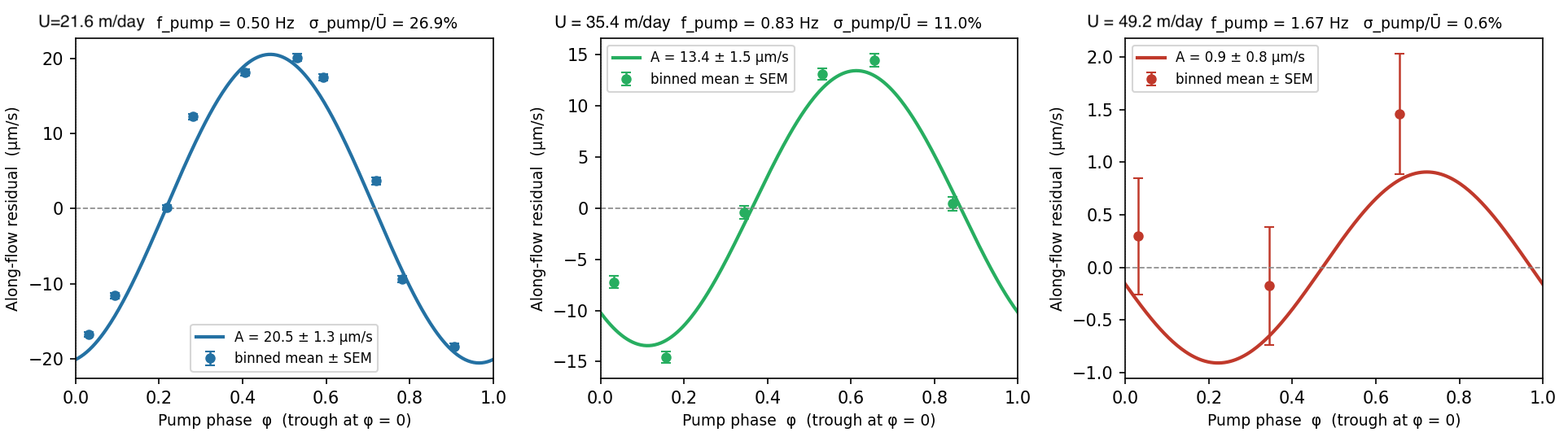}
 \caption{ \textbf{Residual oscillation in flow from peristaltic motion}}
 \label{FigureS2}
 \end{center}
 \end{figure*}
An ensemble PIV is performed to obtain $\mathbf{U}_{\mathrm{mean}}(\mathbf{x})$, classifying passive streamers [see Methods]. For each passive-streamer step at frame $k$, the along-flow residual is computed as $r(k) = \mathbf{v}_{\mathrm{passive}} \cdot \hat{\mathbf{U}}_{\mathrm{mean}} - |\mathbf{U}_{\mathrm{mean}}(\mathbf{x})|$. This is the component of the passive streamer's velocity along the local mean-flow direction, minus the expected mean-flow speed, and is non-zero only because of pump oscillation. The data are phase-folded using $\phi = ((k-k_{\mathrm{trough}}) \bmod T_{\mathrm{frames}}) /
T_{\mathrm{frames}}$, such that $\phi=0$ corresponds to the trough. The data are binned into 16 phase bins and fitted to $A\sin(2\pi\phi+\psi)$, where $\psi$ is a free parameter, avoiding the assumption that the trough occurs exactly at $\phi=0$. Finally, the amplitude $A \pm A_{\mathrm{err}}$ ($\mu\mathrm{m/s}$) is calculated, along with $\sigma_{\mathrm{pump}}/\bar{U} =
A/(\sqrt{2}\,\bar{U}) \times 100$ as a percentage error.
\begin{table}[b]
    \centering
    \caption{Flow conditions and measured velocity statistics.}
    \label{tab:flow_conditions}{
\begin{tabular}{cccccc}
    \hline
    Speed ($\mu$m/s) & $f_{\mathrm{pump}}$ (Hz) & $T$ (s) &
    $A$ ($\mu$m/s) & $\bar{U}$ ($\mu$m/s) &
    $\sigma/\bar{U}$ (RMS, \%) \\
    \hline
    250 & 0.500 & 2.00 & $20.52 \pm 1.26$ & 53.9 & 26.9 \\
    410 & 0.833 & 1.20 & $13.44 \pm 1.48$ & 86.3 & 11.0 \\
    570 & 1.667 & 0.60 & $0.91 \pm 0.83$ & 109.2 & 0.6 \\
    \hline
\end{tabular}
}\end{table}

\section*{II. Stability Analysis of Orientations in Stokesian Weak Advection limit}

The dynamics of a chemotactic swimmer around a moving source consists of two parts:(1) the flow around the source object shapes the chemical field, and (2) the swimmer responds to both ambient flow, its gradients and the chemical field. For simplicity, we assume this field to be a single scalar. We work in a dilute limit, where the dynamics of individual microbes are decoupled from the motion of others. The chemical field $\phi$ produced by the flow $\mathbf{v}$ past stationary source sphere of radius $a$ is governed by the advection diffusion equation
\begin{equation}
 Pe \, (\mathbf{U}\cdot \mathbf{\nabla} \phi)  = \nabla^{2} \phi
 \label{eqn:AD}
\end{equation}
Here, $Pe = U_{\infty}a/D$ is the P\'{e}clet number with diffusivity $D$ and $U_{\infty}\equiv \mathbf{U}(\mathbf{r} \to \infty)$. The dynamics of $\mathbf{U(r,t)}$ is governed by the Navier-Stokes equation,
\begin{equation}
   Re \left(\frac{\vect{\partial U}}{\partial t} \, +\, \vect{U}\cdot\nabla \vect{U}\right) = -\nabla P + \nabla^2 \vect{U}.
   \label{eqn:NS}
\end{equation}
The lengths and times are non-dimensionalized using $a$ and $a/U_{\infty}$, and we consider creeping flow where the pressure scaled by $\mu U_{\infty}/a$ is P, with the dynamic viscosity $\mu$. In this work, we do not solve \eqref{eqn:NS} but instead (1) use the Stokesian solution around a sphere for analytical calculation, and (2) measure $\vect{U}$ using the tracer-like dynamics of non-motile microbes and calculate $\phi$ by solving \eqref{eqn:AD} with Dirichlet boundary condition with a fixed value $\phi_{0}$ at the boundary of the aggregate, as discussed below. The details of the numerical calculation are given in the Methods section.

In the limit of zero Reynolds number $Re \to 0$ of Eq. \eqref{eqn:NS}, we get the flow field
\begin{equation}
\mathbf{U}(r,\theta)
=
\begin{pmatrix}
\displaystyle
\left(1-\frac{3}{2r}+\frac{1}{2r^3}\right)\sin\theta
\\[8pt]
\displaystyle
\left(1-\frac{3}{4r}-\frac{1}{4r^3}\right)\cos\theta
\end{pmatrix},
\label{StokesU}
\end{equation}
and $\nabla \phi$ comes from  the inner solution of the singular perturbation analysis of Eq. \eqref{eqn:AD}. To leading order in P\'eclet number $\phi$ is

\begin{equation}
\phi(r) = \frac{1}{r}
+ Pe\left[
\frac{1}{2}\left(\frac{1}{r}-1\right)
+
\left(
\frac{1}{2}
-\frac{3}{4r}
+\frac{3}{8r^2}
-\frac{1}{8r^3}
\right)\frac{y}{r}
\right],
\label{phiInner}
\end{equation}

 The orientational dynamics, without noise, of a rigid chemotactic swimmer with unit vector $\vect{p} = (\cos \theta, \sin\theta)$ along the axisymmetric direction, responding to the flow gradient $\nabla \vect{v}$ and chemical gradient $\nabla \phi$ generated by a sinking particle is
\begin{equation}
    \dot{\vect{p}} = (\mathbb{I} - \vect{pp})\cdot(\alpha\mathbb{S} + \Omega)\cdot\vect{p} \, + \, \gamma \,\vect{p}\times(\nabla \phi \times\vect{p}),
    \label{EqJefferyChemotaxis}
\end{equation}
where $\alpha$ is the Jeffery's coupling \cite{Jeffery1922} to the symmetric $\mathbb{S} = (\nabla\vect{v} + \nabla\vect{v}^{T})/2$ part of the flow gradient, and $\Omega = (\nabla\vect{v} - \nabla\vect{v}^{T})/2$ is the antisymmetric or vortical part given by,

\begin{equation}
\begin{aligned}
\mathbf{S}(r,\theta)
&=
\begin{pmatrix}
\displaystyle \frac{3(r^2-1)\sin\theta}{2r^4}
&
\displaystyle \frac{3\cos\theta}{4r^4}
\\[8pt]
\displaystyle \frac{3\cos\theta}{4r^4}
&
\displaystyle -\frac{3(r^2-1)\sin\theta}{4r^4}
\end{pmatrix},
\qquad
&
\boldsymbol{\Omega}(r,\theta)
&=
\begin{pmatrix}
0 & -\dfrac{3\cos\theta}{4r^2}
\\[8pt]
\dfrac{3\cos\theta}{4r^2} & 0
\end{pmatrix}.
\end{aligned}
\label{gradU}
\end{equation}

The term with chemotactic coupling $\gamma$ becomes
\begin{align}
\vect{p}\times(\nabla \phi \times\vect{p}) &= \begin{pmatrix} \sin^{2}\theta \, \nabla_{x} \phi - \sin \theta \cos \theta \nabla_{y} \phi \\ \cos^{2} \theta \nabla_{y} \phi - \sin\theta \cos\theta \nabla_{x} \phi \end{pmatrix} \nonumber \\
&= \begin{pmatrix} \vect{p}_{y}^{2} & - \vect{p}_{x} \vect{p}_{y} \\ - \vect{p}_{x} \vect{p}_{y} & \vect{p}_{x}^{2}\end{pmatrix} \begin{pmatrix} \nabla_{x} \phi \\ \nabla_{y} \phi \end{pmatrix} \nonumber \\
& = (\mathbb{I} - \vect{pp})\cdot \vect{\nabla} \phi \label{EqChemSimplify}
\end{align}

where we have used $\vect{p}_{x}^{2} + \vect{p}_{y}^{2} = 1$.  From equations \eqref{EqJefferyChemotaxis} \& \eqref{EqChemSimplify}
\begin{equation}
    \dot{\vect{p}} = (\mathbb{I} - \vect{pp})\cdot [ (\alpha\mathbb{S} + \Omega) \cdot\vect{p} \, + \, \gamma \, \vect{\nabla} \phi ] 
    \label{EqSteadyp_supp}
\end{equation}
yields the steady state solution of the swimmer orientation $\vect{p}$. Since $(\mathbb{I} - \vect{pp})$ is the projection perpendicular to $\vect{p}$, \eqref{EqSteadyp_supp} is zero when 
\begin{equation}
    (\alpha\mathbb{S} + \Omega) \cdot\vect{p} \, + \, \gamma \, \vect{\nabla} \phi = \lambda \vect{p}.
\end{equation}
that is, when $\mathbb{I} - \vect{pp}$ operates on a vector parallel to $\vect{p}$. $\gamma=0$ case presents an eigenvalue problem which we solve in Mathematica for $\alpha=1$ (needle-like particles) that responds most strongly to $\mathbf{S}$. The complex eigen-values produces the vorticity-dominated region of instability shown in Fig. 3(d).

The $\gamma \neq 0$ case is more interesting. Equation (14) gives the steady-state solution for swimmer orientation
\begin{equation}
    \vect{p} =  [\lambda \mathbb{I} - (\alpha\mathbb{S} + \Omega) ]^{-1} \cdot \gamma\nabla \phi
\end{equation}
which is readily solved analytically for $\alpha=0$ (spherical swimmers) in terms of the local vorticity $\omega$, producing the chemotactic horizon. In the manuscript we perform the stability analysis for a complex shaped aggregate using measured values of $U, \mathbf{S}$ and $\mathbf{\Omega}$, and the numerically inferred $\nabla \phi$, which produces the chemotactic horizons.

\section*{III. Chemotactic turning response through an effective inertial dynamics}
We begin with first order equations of motion for swimmer position $\vect{X}$, orientation $\vect{p}$ and in intrinsic swim speed $v_0$, in the deterministic sector of the full dynamics,
\begin{equation}
\dot{\mathbf{X}}  =  \mathbf{U}|_{\bf r=X} + v_{0} 
{\mathbf{p}} \equiv \vect{v},   
\label{eqnX5}
\end{equation}

\begin{align}
\dot{\mathbf{p}} & =  [\mathbb{I} - \vect{p}\vect{p}]\cdot(\alpha \bsf{S} + \vect{\Omega})\cdot \mathbf{p} \, + \, v_{0} \gamma \mathbf{p} \times (\nabla \bar{\phi} \times \mathbf{p})
\label{eqnS6}
\end{align}

Here, $\bsf{S}$ and $\vect{\Omega}$ are the symmetric and the antisymmetric part of $\vect{\nabla U}$, with $\vect{p}$ coupling to $\bsf{S}$ through a shape dependent coupling $\alpha$ \cite{Jeffery1922}. The chemotactic term in \eqref{eqnS6} $\, v_{0}\gamma \mathbf{p} \times (\nabla \bar{\phi} \times \mathbf{p})$ coupling $\vect{p}$ to the gradients of $\bar{\phi} \equiv \phi/\phi_{0}$, where $\phi_{0}$ is the concentration at the source, is constructed on general symmetry grounds, retaining leading orders in the gradient expansion. The projection operator $\mathbb{I} - \vect{p}\vect{p}$ in \eqref{eqnS6} ensures unit magnitude of $\vect{p}$. Since the skew symmetric tensor $\vect{\Omega}$ is self-projecting $\vect{\Omega} \cdot \vect{p}$ does not need a projection operator in front of it.

\begin{figure*}[t]
 \begin{center}
 \includegraphics[width=17.5 cm]{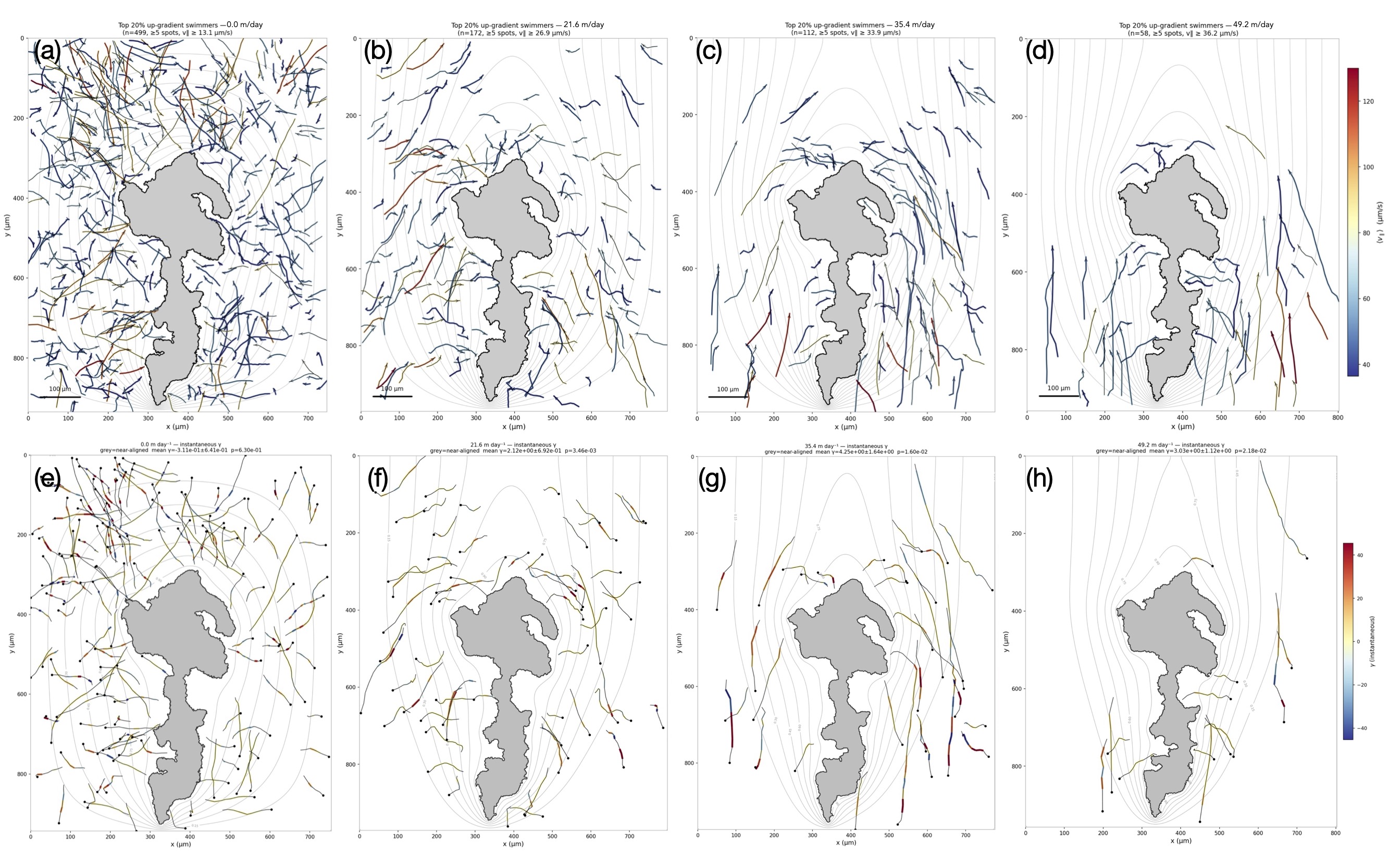}
 \caption{ \textbf{Chemotactic trajectories}: trajectory from the top 20\% of chemotactic attractors, selected based on the distribution of $\langle \mathrm{v}_{||}\rangle_{\text{track}}$, and colored based of that value, for varying flows: (a) 0 m/day (b) 21.6 m/day (c) 35.4 m/day (d) 49.2 m/day. The chemotactic turning response $\gamma$ plotted on the corresponding tracks (e)-(h). }
 \label{FigureS3}
 \end{center}
 \end{figure*}

Along similar lines to the derivation of effective inertial dynamics in \cite{ChajwaCaus2026}, we take the time derivative of \eqref{eqnX5}, substituting $\dot{\vect{p}}$ using \eqref{eqnS6} and rewrite $\vect{p} = (\vect{v} - \vect{U})/v_0$ using \eqref{eqnX5}, which gives
\begin{align}
    \frac{d\vect{v}}{dt} &=  \frac{\partial \vect{U}}{\partial t} + \vect{v}\cdot \vect{\nabla U} + \vect{\Omega} \cdot (\vect{v} - \vect{U}) + \gamma v_{0} ^{2} \vect{\nabla} \bar{\phi} - \gamma \left(\vect{v} - \vect{U}\right) \left(\vect{v} - \vect{U}\right) \cdot \vect{\nabla} \bar{\phi} 
\end{align}
In steady flows $\partial \vect{U}/\partial t = 0$. Decomposing $\nabla \vect{U}$ into symmetric and antisymmetric components give,
\begin{align}
    \frac{d\vect{v}}{dt} &=  \vect{v}\cdot \bsf{S} + \vect{v}\cdot\vect{\Omega} + \vect{\Omega} \cdot \vect{v} - \vect{\Omega} \cdot\vect{U} + \gamma v_{0} ^{2} \vect{\nabla} \bar{\phi} - \gamma \left(\vect{v} - \vect{U}\right) \left(\vect{v} - \vect{U}\right) \cdot \vect{\nabla} \bar{\phi}  \nonumber \\
    &= \vect{v}\cdot \bsf{S} - \vect{\Omega} \cdot\vect{U} + \gamma v_{0} ^{2} \vect{\nabla} \bar{\phi} - \gamma \left(\vect{v} - \vect{U}\right) \left(\vect{v} - \vect{U}\right) \cdot \vect{\nabla} \bar{\phi}
\end{align}
where we have substituted $\vect{v} \cdot \vect{\Omega} + \vect{\Omega}\cdot \vect{v} = \vect{\Omega}^{\mathsf{T}} \cdot \vect{v} + \vect{\Omega}\cdot \vect{v} = 0$, since $\vect{\Omega}$ is skew-symmetric, that is $\vect{\Omega}^{\mathsf{T}} = -\vect{\Omega}$. Taking the dot product of both sides with $\nabla \phi$ and solving for $\gamma$ in terms of $\dot{\vect{v}}, \vect{v}, \vect{U}, \bsf{S}, \vect{\Omega},$ and $\nabla \phi$ gives 
\begin{equation}
    \gamma = \frac{(\dot{\vect{v}} \, - \,  \bsf{S} \cdot \vect{v} \, + \vect{\Omega} \cdot \vect{U} ) \cdot \nabla \bar{\phi}}{v_{0}^{2} \lVert \nabla \bar{\phi} \rVert^{2} \sin^{2} \delta } 
    \label{eqnSgamma}
\end{equation}
where $\delta$ is the angle between $\vect{p}$ and $\nabla \phi$. Since \eqref{eqnSgamma} diverges for $\delta \to 0$, we only calculate chemotactic turning response $\gamma$ when $\delta \neq 0$, that is when the swimmer is not aligned with $\nabla \phi$ ($\sin^{2}\delta > 0.05$), for highly chemotactic swimmers with at least 8 spots in the track [see Fig. \ref{FigureS3}(e)-(h)].

\section*{Dynamical phase diagram}
The outer solution to Eqn. \eqref{eqn:AD} to leading orders in Pe, in plane polar coordinates is
\begin{equation}
    \phi(r,\theta) = \frac{1}{r}\exp{\left[-\frac{Pe}{2} r(1-\cos\theta)\right]}
    \label{eqn_outer}
\end{equation}
where $\theta$ is measured from the positive $z$-axis. From this we calculate $\lVert\nabla \phi\rVert$. The various horizons are calculated as follows:
(i) Sensory Horizon (Berg-Purcell): $\kappa \lVert \nabla \phi\rVert/\sqrt{\phi} =1$, where $\kappa = v_0 T^{3/2}\sqrt{4 \pi D \phi_0/b}$. The parameters used to plot the sensory horizon in Fig. \ref{Phase_diagram} is given in Table \ref{sensory_horizon_parameters}, which yields $\kappa =42.6$. (ii) Chemotactic horizon: $2v_0 \gamma \lVert \nabla \phi\rVert/\omega =1$, with vorticity from Eqn. \eqref{gradU} and measured $\gamma = 2.88$. (iii) Chasing horizon, the level curve $u_z = v_0$ with $u_z$ from Eqn. \eqref{StokesU}.
\begin{table}[ht]
\centering
\caption{Parameters used to plot the sensory horizon.}
\label{sensory_horizon_parameters}
\begin{tabular}{lll}
\hline
Parameter & Symbol & Value \\
\hline
Aggregate radius & $a$ & $100~\mu\mathrm{m}$ \\
Aggregate sinking speed & $U$ & $50~\mu\mathrm{m}\,\mathrm{s}^{-1}$ \\
Microbial swimming speed & $v_0$ & $40~\mu\mathrm{m}\,\mathrm{s}^{-1}$ \\
Molecular diffusivity & $D$ & $2000~\mu\mathrm{m}^2\,\mathrm{s}^{-1}$ \\
Effective cell size & $b$ & $1~\mu\mathrm{m}$ \\
Sensing integration time & $T$ & $1~\mathrm{s}$ \\
Effective surface concentration & $\phi_0$ & $0.75~\mathrm{nM}$ \\
\hline
\end{tabular}
\end{table}

 \section*{V. Solving 1D Transport Equation for sinking and degrading marine snow}
We solve Eqn. \eqref{Transporteqn1} for the initial log-normal distribution of speeds located at $z_0 = 10\mathrm{m}$ fitted on measured speeds from Fig. \ref{Figure3}(a).
\begin{equation}
p(v_0) = \frac{1}{v_0 \sigma \sqrt{2\pi}}{\exp\left[-\dfrac{(\ln v_0 - \mu)^2}{2\sigma^2}\right]},
\qquad \mu = 3.90,\quad \sigma = 1.25.
\end{equation}
\subsection*{Particle dynamics}
We first solve for the characteristics with initial condition $(z_0, v_0)$, with $v_0 < v_c$
\begin{equation}
\begin{aligned}
\frac{d\mathrm{z}}{dt} &= v
\qquad &
\frac{d\mathrm{v}}{dt} &= -R_{0}\cos(qv-\theta_0)\, v
\end{aligned}
\end{equation}
These equations gives the phase space (z,v) dynamics
\begin{equation}
    \frac{dz}{dv}  = -\frac{1}{R_{0}\cos(qv-\theta_0)},
\end{equation}
with change of coordinate $\theta = \pi(v-v_m)/2(v_c - v_m)$which gives the solution
\begin{equation}
z - Z_0 = A \int_{\theta}^{\theta_0} \frac{d\theta'}{\cos\theta'}
= A \left[ \ln\left(\sec\theta' + \tan\theta'\right) \right]_{\theta}^{\theta_0},
\end{equation}
where $A \equiv 2\Delta/(\pi R_0) = 311.9\,\mathrm{m}$, with $\Delta \equiv v_c - v_m$. The solution can be written in terms of $s = \ln\left(\sec\theta + \tan\theta\right)$,
\begin{equation}
    z = z_0 + A(s_0 - s)
\end{equation}
 where $s_0$ is the value of $s$ at particle's initial speed $v_0$. The speed is recovered by inverting the definition,
 \begin{equation}
v = G(s) = v_m + \frac{2\Delta}{\pi}\arctan(\sinh s),
\qquad
\text{and} \,\, R = \frac{R_0}{\cosh s}.
\end{equation}
The particle with initial speed $v_0$ has a closed form velocity $v(z) = G\left(s_0 - \frac{z-Z_0}{A}\right)$ at depth $z$. These curves plotted in the phase portrait in Fig. \ref{Figure4}(b). The particle stalls (v = 0) at a finite depth giving the line of critical points in Fig. \ref{Figure4}(b). Note that $R(0) = 0.94 R_0 \neq 0$, so the curve meets $v = 0$ at a slant. Although we know the phase curves analytically, at what time the particle reaches depth $z$ is still not determined, and requires a numerical quadrature in the solution of
\begin{equation}
T(s) = A \int \frac{ds'}{G(s')}.
\end{equation}
which we tabulate once on a fine grid. Then $t = T(s_0) - T(s)$. To find where particle with initial speed $v_0$ is at time $t$, we solve solve $T(s) = T(s_0) - t$ for a given $s$ by linear interpolation. That gives its depth $z = z_0 + A(s_0 - s)$ and speed $v = G(s)$.

\begin{figure*}[t]
 \begin{center}
 \includegraphics[width=12 cm]{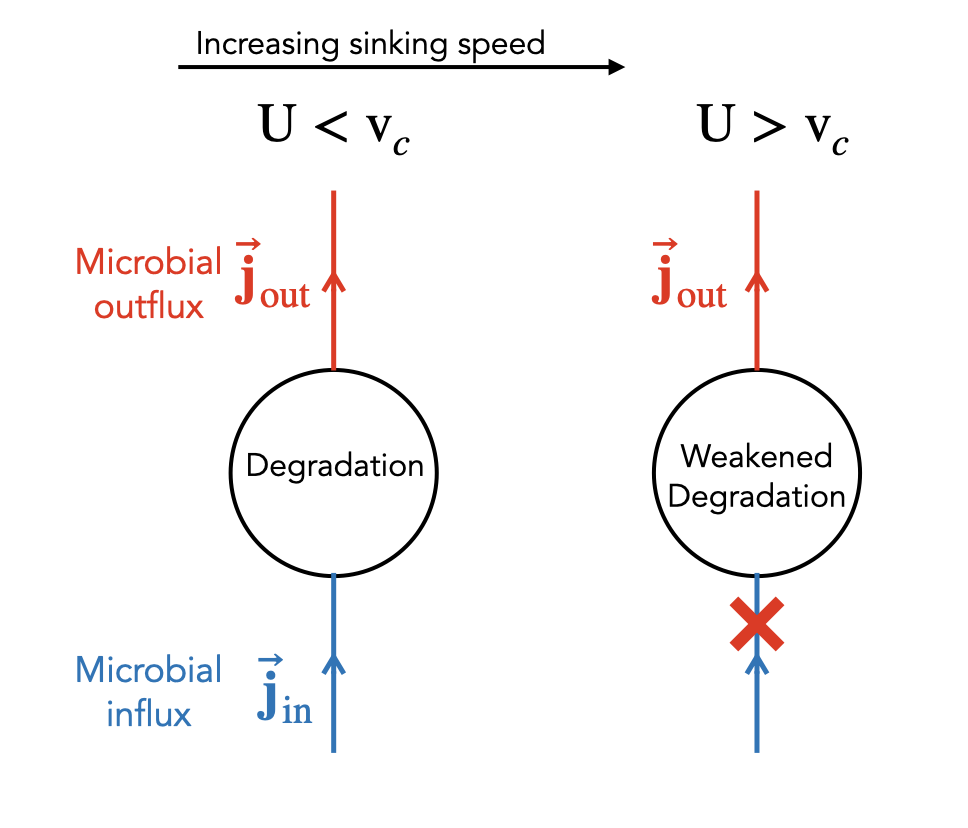}
 \caption{ \textbf{Microbial flux-mediated degradation model}: As the marine snow sinks through vertically stratified microbial niches in the ocean, microbe-marine snow encounters produces an influx and outflux of microbes executing degradation. If the marine snow particle sinks faster than a critical sinking speed, $\mathrm{U}=\mathrm{v_c}$ the influx $\vec{\vect{j}}_{\mathrm{in}}$ drops to zero, collapsing the steady state to produce weakened degradation. This implies that the faster sinking marine snow particles can effectively evade colonization by new microbes and hence lead to reduce degrdation. Mathematically the assumption is that degradation is a function of sinking speed $v$, following the same profile as the encounter rate. }
 \label{FigureS4}
 \end{center}
 \end{figure*}

\subsection*{From particles to field}

\textit{Density field $\rho(z,t)$}: The initial number-density distribution is $\delta(z_0)$, which should be thought of as corresponding to a sheet of particles at an initial depth $z_0$ in the ocean. Note that, the 3D system is assumed isotropic in the subspace perpendicular to gravity, and our 1D model projects the dynamics onto a vertical line, where this sheet is mapped onto a point $z_0$). Particle number conservation implies
\begin{equation}
\rho(z,t)\,dz = p(v_0)\,dv_0
\quad\Longrightarrow\quad
\rho(z,t) = \frac{p(v_0)}{\left.\dfrac{\partial z}{\partial v_0}\right|_t}.
\end{equation}
where ${\partial z}/{\partial v_0}$ is the Jacobian of the map $z = z_0 + A[s_0(v) - s(v)]$ evaluated at a fixed time $t$, 
\begin{equation}
\frac{\partial z}{\partial v_0}
= A\frac{ds_0}{dv_0}\left(1-\frac{ds}{ds_0}\right).
\end{equation}
Differentiating $t=T(s_0)-T(s)$ with respect to $s_0$ at fixed $t$, and using $T'(s)=A/G(s)$, gives
\begin{equation}
\frac{ds}{ds_0}
= \frac{G(s)}{G(s_0)}
= \frac{v}{v_0}.
\end{equation}
Differentiating $v=G(s)$ gives ${dv}/{ds}
= {2\Delta/\pi}/{\cosh s}$, from which

\begin{figure*}[t]
 \begin{center}
 \includegraphics[width=10 cm]{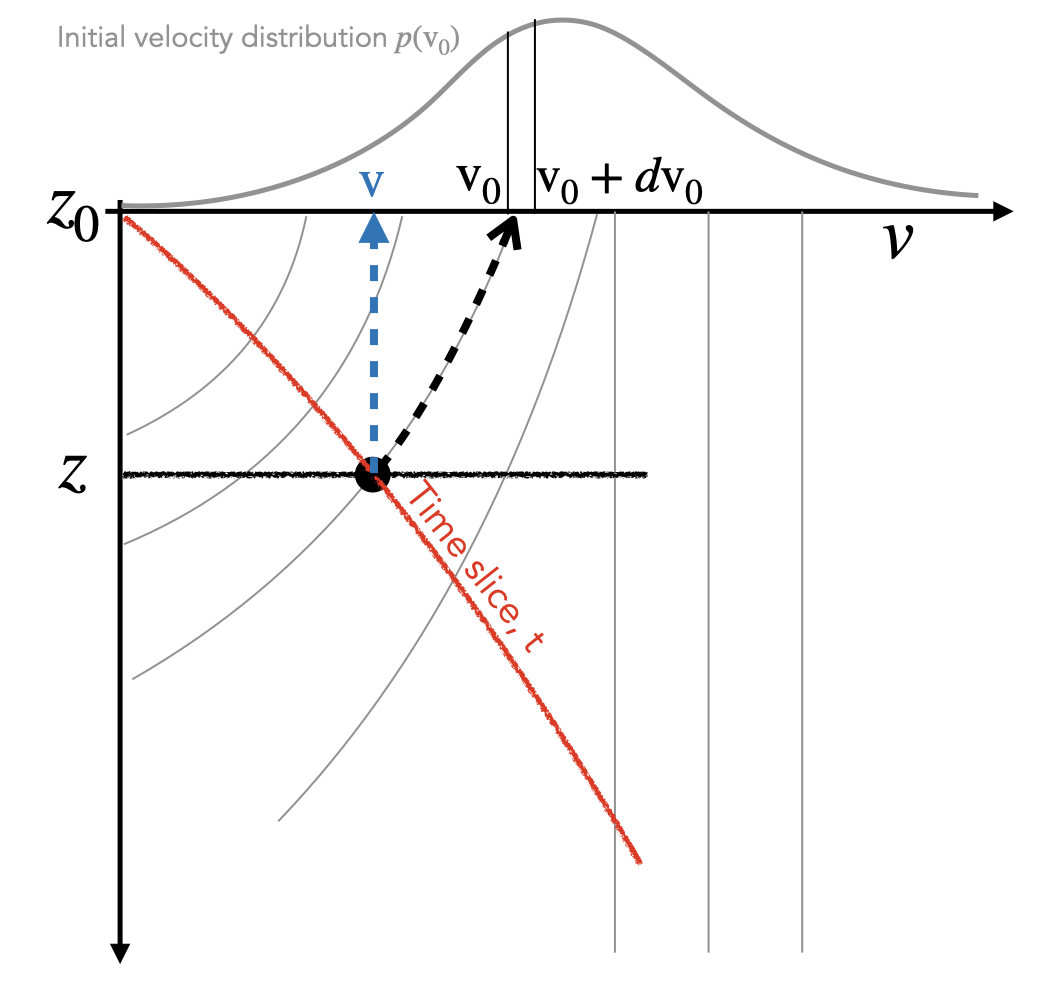}
 \caption{ \textbf{Geometric picture of the solution scheme for the density field:} To calculate density $\rho(z,t)$ at point (z,t) for in the overdamped region of the phase-space, take a constant time slice at 't' of the phase flow from T(s), and calculate where it intersects with constant $z$ curve which gives $(z,v)$. From this point of intersection follow the phase curve backwards in time to get $\mathrm{v}_0$, which gives $R(\mathrm{v_0})$ and $p(\mathrm{v_0})$, yielding $\rho(z,t)$ from (34).}
 \label{FigureS5}
 \end{center}
 \end{figure*}

\begin{equation}
\frac{ds_0}{dv_0}
= \frac{1}{A R(v_0)}.
\end{equation}
From (30), (31) \& (33)  we get
\begin{equation}
\frac{\partial z}{\partial v_0}
= \frac{1}{R(v_0)}
\left(1-\frac{v}{v_0}\right).
\end{equation}
which yields the density field in a closed form:
\begin{equation}
    \rho(z,t) = \frac{p(v_0)\, R(v_0)}{\left(1- \frac{v}{v_0}\right)}
\end{equation}
valid for $v \neq v_0$, that is $t>0$. For fast particles ($v_0 > v_c$), $\partial z/\partial v_0 = t$, which gives the solution for ballistic region in the phase-space $\rho(z,t)= p((z-z_0)/t)/t$.

\textit{Velocity field $v(z, t)$}: At time $t$, we plot each particle’s position $z(v_0, t)$ against its speed $v(v_0, t)$ and acquire the curve $v(z)$. We find that $z$ increases monotonically with $v_0$. Below depth $z_0 + v_c t$ only fast particles are present, and there $v = (z - z_0)/t$.

\end{document}